\documentclass[12pt, draftclsnofoot, onecolumn]{IEEEtran}

\usepackage{cite}

\usepackage{amsmath}
\usepackage{amssymb}

\usepackage{algorithm}
\usepackage{algorithmic}
\algsetup{linenosize=\footnotesize}

\usepackage{hyperref}

\DeclareMathOperator*{\argmin}{arg\,min}

\usepackage{booktabs}
\usepackage{setspace}

\usepackage[acronyms,nonumberlist,nopostdot,nomain,nogroupskip]{glossaries}

\newignoredglossary{ignored}

\newacronym[type=ignored]{vanet}{VANET}{Vehicular Ad Hoc Network}
\newacronym[type=ignored]{ukf}{UKF}{Unscented Kalman Filter}
\newacronym[type=ignored]{cdf}{CDF}{Cumulative Distribution Function}
\newacronym[type=ignored]{iid}{IID}{Independent and Identically Distributed}
\newacronym[type=ignored]{ctra}{CTRA}{Constant Turn Rate and Acceleration}
\newacronym[type=ignored]{kf}{KF}{Kalman Filter}
\newacronym[type=ignored]{pf}{PF}{Particle Filter}
\newacronym[type=ignored]{hmm}{HMM}{Hidden Markov Model}
\newacronym[type=ignored]{its}{ITS}{Intelligent Transport System}
\newacronym[type=ignored]{ml}{ML}{Machine Learning}
\newacronym[type=ignored]{svm}{SVM}{Support Vector Machine}
\newacronym[type=ignored]{nn}{NN}{Neural Network}
\newacronym[type=ignored]{gps}{GPS}{Global Positioning System}
\newacronym[type=ignored]{dr}{DR}{Dead Reckoning}
\newacronym[type=ignored]{bf}{BF}{Bayesian Filtering}
\newacronym[type=ignored]{rbf}{RBF}{Radial Basis Function}
\newacronym[type=ignored]{cits}{C-ITS}{Connected and Intelligent Transportation Systems}
\newacronym[type=ignored]{sumo}{SUMO}{Simulation of Urban MObility}
\newacronym[type=ignored]{dsrc}{DSRC}{Dedicated Short Range Communication}
\newacronym[type=ignored]{csmaca}{CSMA/CA}{Carrier Sense Multiple Access with Collision Avoidance}
\newacronym[type=ignored]{qoi}{QoI}{Quality of Information}
\newacronym[type=ignored]{aoi}{AoI}{Age of Information}
\newacronym[type=ignored]{mac}{MAC}{Medium Access Control}
\newacronym[type=ignored]{pdf}{PDF}{Probability Density Function}
\newacronym[type=ignored]{cbrl}{$CBR_{local}$}{Local Channel Busy Ratio}
\newacronym[type=ignored]{cbrv}{$CBR_{vehicle}$}{Vehicle Channel Busy Ratio}
\newacronym[type=ignored]{cbrt}{$CBR_{target}$}{Target Channel Busy Ratio}
\newacronym[type=ignored]{cav}{CAV}{Connected and Autonomous Vehicle}
\newacronym[type=ignored]{v2v}{V2V}{Vehicle-To-Vehicle}
\newacronym[type=ignored]{tdma}{TDMA}{Time Division Medium Access}
\newacronym[type=ignored]{wave}{WAVE}{Wireless Access in Vehicular Environment}
\newacronym[type=ignored]{rsu}{RSUs}{Roadside Units}
\newacronym[type=ignored]{pb}{PB}{Periodic Broadcasting}
\newacronym[type=ignored]{etb}{ETB}{Error Threshold Broadcasting}
\newacronym[type=ignored]{cscc}{CSCC}{Channel Sensing Congestion Control}
\newacronym[type=ignored]{nacc}{NACC}{Neighbor Aware Congestion Control}
\newacronym[type=ignored]{ppp}{PPP}{Poisson Point Process}

\usepackage{graphicx}
\usepackage{placeins}
\usepackage{float}
\usepackage{caption}
\usepackage{subcaption}

\usepackage{color}

\usepackage[font=footnotesize]{caption}
\usepackage[font=scriptsize]{subcaption}

\begin{document}

\makeatletter
\renewcommand{\ALG@name}{\footnotesize Algorithm}
\makeatother

\title{An Adaptive Broadcasting Strategy for Efficient Dynamic Mapping in Vehicular Networks}

\author{{{Federico Mason}, {Marco Giordani},~\IEEEmembership{Student Member, IEEE}, \\ {Federico Chiariotti},~\IEEEmembership{Member, IEEE}, {Andrea Zanella},~\IEEEmembership{Senior Member, IEEE}, {Michele Zorzi},~\IEEEmembership{Fellow, IEEE}}

\thanks{The authors are with the Department of Information Engineering, University of Padova, Padova, Italy, Email: \{masonfed, giordani, chiariot, zanella, zorzi\}@dei.unipd.it.

A preliminary version of this paper that did not consider congestion control was
presented at the \emph{25th European Wireless Conference (EW)}, May 2019 \cite{mason2018feasibility}.
}}

\maketitle

\vspace{-50pt}
\begin{abstract}
In this work, we face the issue of achieving an efficient dynamic mapping in vehicular networking scenarios, i.e., to obtain an accurate estimate of the positions and trajectories of connected vehicles in a certain area.
State of the art solutions are based on the periodic broadcasting of the position information of the network nodes, with an inter-transmission period set by a congestion control scheme.
However, the movements and maneuvers of vehicles can often be erratic, making transmitted data inaccurate or downright misleading. 
To address this problem, we propose to adopt a dynamic transmission scheme based on the actual positioning error, sending new data when the estimate passes a preset error threshold. 
Furthermore, the proposed method adapts the error threshold to the operational context according to a congestion control algorithm that limits the collision probability among broadcast packet transmissions.
This threshold-based strategy can reduce the network load by avoiding the transmission of redundant messages, and is shown to improve the overall positioning accuracy by more than 20\% in realistic urban~scenarios.

\end{abstract}

\vspace{-10pt}

\IEEEpeerreviewmaketitle

\section{Introduction}

\label{sec:intro}
In recent years, there has been a growing interest in vehicular communications, which have rapidly emerged as a means to support safe and efficient transportation systems through inter-vehicular networking~\cite{hartenstein2008tutorial}. 
From a safety perspective, vehicular networks can mitigate the severity of traffic accidents by notifying the vehicles about dangerous situations in their surroundings, including bad road conditions and approaching emergency vehicles~\cite{hossain2010vehicular}.
Moreover, they can also support other services, ranging from real-time multimedia streaming to interactive gaming and web browsing~\cite{amadeo2012enhancing}.

As the automotive industry looks towards \gls{cits} to support safety-critical applications~\cite{lu2014connected}, research has focused on the design of novel \gls{cits} architectures that guarantee the timely and accurate positioning of vehicles~\cite{Boukerche2008vehicular}.
Positioning is typically provided by on-board \gls{gps} receivers, which may not always have the required accuracy~\cite{alam2013cooperative}.
For this reason, \emph{data fusion} techniques have also been considered in \gls{cits} by combining several positioning strategies (including, but not limited to, dead reckoning, map matching, and camera image processing) into a single solution that is more robust and precise than any individual approach~\cite{Balico:2018}.
However, urban vehicular mobility scenarios may involve rapid dynamics and unpredictable changes in the network topology~\cite{yousefi2006vehicular}, which would require position updates to be disseminated as timely as possible, ideally at the very same instant they are generated.
At the same time, broadcasts over bandwidth-limited communication channels are prone to packet collisions~\cite{joo2018wireless}.

In this scenario, the traditional approach is to have each vehicle broadcast periodic updates with its positioning information. 
However, the intrinsically variable topology of vehicular networks might make periodic broadcasting strategies inefficient: long inter-transmission intervals may prevent the timely dissemination of positioning information in safety-critical situations, while very frequent broadcasting may overload the wireless medium with useless data and increase the number of packet collisions. Congestion avoidance mechanisms have thus been proposed in the literature to regulate information distribution as a function of the network load~\cite{wischhof2005congestion}. 
These techniques dynamically adapt to the number of neighboring vehicles~\cite{caizzone2005power} or assign priorities to vehicles based on their operating conditions~\cite{huang2009analysis}, but usually disregard the level of positioning accuracy that is finally achieved.


To solve these issues, more sophisticated information distribution solutions that explicitly consider the \emph{\gls{qoi}}~\cite{bisdikian2013quality} have been investigated.
These techniques consider the value of the possible positioning information updates, only broadcasting those that maximize the utility for the target applications\cite{higuchi2019value,giordani201investigating}. 
The \gls{qoi} assessment process should be computationally efficient, so that it can be executed in real-time even with the limited on-board computational resources of mid-range and budget car models.

Following this rationale, in this paper we face the challenge of ensuring accurate position estimation of vehicles while minimizing the network load in a cost-effective way.
The main novelty of our work consists in the following points:

\begin{itemize}
	
	\item We find a mathematical expression of the packet collision probability as a function of the vehicular traffic density.
	
	\item Based on the above-mentioned relation, we design a new congestion control mechanism that exploits network topology information to reduce the packet collision probability.
	Compared with traditional channel-based congestion control, our solution can better adapt to fluctuating conditions of the environment.
	
	\item We design a \emph{threshold-based} broadcasting algorithm that \emph{(i)} estimates the positioning error of the vehicle and its neighbors within communication range, based on a purely predictive \gls{ukf}, and \emph{(ii)} makes vehicles distribute state information messages if the estimated error is above a predefined threshold.
	
	\item We investigate the performance of the proposed scheme using realistic mobility traces generated by \gls{sumo}~\cite{krajzewicz2012recent}, an open road traffic simulator designed to handle and model the traffic of large road networks. 
	
	
	
\end{itemize}
The performance of our approach is compared with a baseline \emph{periodic broadcasting} solution that instructs vehicles to broadcast state information at regular intervals. Simulation results show that the proposed algorithm, in spite of its simplicity, can reduce the average position estimation error by more than 10\%, and its 95th percentile by more than 20\%, compared to the periodic broadcast approach.

The remainder of this paper is organized as follows. In Sec.~\ref{sec:related_work} we present a selection of the most relevant related work. In Sec.~\ref{sec:model} we introduce our system model. In Secs.~\ref{sec:comm_strategy} and~\ref{sec:cong_con} we describe our broadcasting strategies and congestion control mechanisms, respectively, and derive the expression for the packet collision probability as a function of the vehicular traffic density. In Sec.~\ref{sec:analysis} we validate our theoretical analysis through simulations and present our main findings and results.
Finally, in Sec.~\ref{sec:conclusions} we provide conclusions and suggestions for future work.

\section{Related Work}
\label{sec:related_work}
In a \gls{cits} scenario, vehicles are equipped with on-board sensors, which are used to gather data about the surrounding environment.
These data are then disseminated within the network through wireless technologies.
Hence, each vehicle can build a dynamic mapping of the other vehicles in its surroundings~\cite{Boukerche2008vehicular}. 
The performance of such a tracking system is highly dependent on the cooperation among vehicles. 
An example of a tracking framework for \gls{cits} is given in \cite{Ramos:2012}: most research in this area is based on similar architectures, but with different vehicle mobility and data processing schemes. An analysis of the main mobility models used in vehicle tracking is given in \cite{Schubert:2008}.
For what concerns data processing, the most common choice is adopting a \gls{bf} approach, typically based on the \gls{kf} \cite{Kalman:1960}, the \gls{ukf} \cite{Wan:2000} or the \gls{pf} \cite{Moral:1996}.
A tracking framework based on the \gls{ukf} and the \gls{ctra} motion model is presented in\cite{Lytrivis:2011}. 
In \cite{Peker:2011}, route information and digital map data are jointly processed by a particle filter algorithm.
In \cite{Akabane:2017}, position forecasting is achieved by using a Hidden Markov Model \cite{Poritz:1988} and the Viterbi algorithm \cite{Viterbi:1967}. 
We highlight that, in all the \gls{bf}-based architectures, the performance greatly depends on the filter settings, e.g., the process and estimation noise covariances, which must be known \emph{a priori} \cite{Roth:2014}. 

Conventional tracking approaches mainly focus on the real-time estimation of the target state.
However, most advanced \gls{cits} applications also require a prediction of vehicles' future trajectories. Long-term forecasting can be achieved by simply applying the predictive step of a \gls{bf} filter to the last available state estimate.
However, this solution is very sensitive to imperfections of the motion model: to overcome this issue, more sophisticated approaches have been proposed in the literature.
In \cite{Kang:2017}, the output of a \gls{kf} is used to perform a parametric interpolation of the future path of the target vehicle.
In \cite{King:2018}, dead reckoning is used to improve the performance of packet forwarding in a highway scenario.
Another possibility consists of describing vehicle position prediction as a time series forecasting problem \cite{Balico:2018}.
Hence, \gls{ml} techniques can improve target state estimation over a large time horizon: in \cite{Wang:2015}, Support Vector Machines are used to forecast vehicle trajectories, allowing the estimation of target positions when the \gls{gps} signal is not available.
In \cite{Park:2011}, a neural network is trained with historical traffic data and then used to predict vehicles' speeds. 
Although the \gls{ml} approach generally guarantees high performance, it requires a large amount of data for the initial training and suffers from significant computational complexity.
\gls{ml} techniques are often combined with \gls{bf} algorithms: in \cite{Deo:2018}, the authors present a system that makes use of a Hidden markov module to estimate vehicle maneuvers and a Support Vector Machine to predict future vehicle trajectories.
In \cite{Hermes:2009}, a Radial Basis Function classifier is used to compute the inner parameters of a particle filter, which estimates the long-term motion of the target.
In \cite{Houenou:2013}, the results of a maneuver recognition system are combined together with the output of a tracking system based on the \gls{ctra} motion model. 

Regardless of the complexity of the tracking framework, the overall system performance degrades if onboard sensor measurements are not sufficiently accurate.
Users can share local information to compensate the low quality of the input data: \gls{cits} nodes periodically broadcast their own system state by using the \gls{dsrc} technology \cite{li2010overview} and the \gls{wave} standard \cite{Jiang:2008}. However, the random channel access scheme may cause congestion in scenarios with a sufficiently high vehicular density: consequently, the transmitted information may be lost because of packet collisions.
Defining novel congestion control schemes, which suit the characteristics of modern vehicular networks, is a problem of interest. 
Over the years, many researchers have proposed different \gls{mac} strategies that adapt inter-vehicle communications to channel conditions. 
In \cite{bansal:2013}, the authors present a rate-adaption strategy that ensures channel stability through vehicular networking.
The convergence of the proposed algorithm is theoretically proved and guidance for the choice of the algorithm parameters is provided. 
In \cite{Omar:2013}, the \textit{hidden terminal} problem is avoided by adopting a time-slotted structure and a Time Division Multiple Access scheme.
In particular, each vehicle is assigned a dedicated timeslot for each frame, during which it can alert its neighbors about its future transmissions. 
In \cite{Taherkhani:2016}, the authors focus on improving congestion control in road intersections by using a locally-distributed strategy based on \gls{ml}. Dedicated road infrastructures have the task of deleting redundant communications and assigning specific \gls{csmaca} parameter settings to different clusters of transmission requests. 

A solution for reducing the channel occupancy is to select the optimal transmission strategy as a function of the instantaneous positioning error of nearby vehicles.
The authors in~\cite{Rezaei:2010}, for example, propose for the first time a broadcasting strategy in which each vehicle triggers new transmissions whenever the estimates of its neighbors' errors are above a predetermined threshold.
However, such analysis is provided only for specific case scenarios and the framework that predicts future vehicles' states is quite obsolete with respect to current vehicular tracking techniques. 
Similarly, in~\cite{Huang:2010}, the transmission rate by which new information is disseminated within the network is regulated according to both the positioning error and the estimated number of packet collisions. 
Nevertheless, the authors assume that vehicles are always aware of the number of packets lost during each timeslot. This may not be always true in vehicular scenarios where most packet collisions are caused by the hidden terminal problem and, thereby, cannot be directly sensed by other nodes. 
In \cite{Fallah:2011}, the authors analyze the inter-vehicular communication dynamics that cause the hidden terminal problem.
In the same work, the limitations of the \gls{csmaca} protocol are addressed by varying the vehicle communication range according to the channel occupancy. 
However, the validity of the approach is proven only in a highway scenario and cannot be generalized to more complex and unpredictable environments.
The adaptation of the broadcasting period according to the channel conditions, as well as the implementation of high performance tracking frameworks, plays a key role in the future development of vehicular networks.
In this perspective, our work aims at designing new communication strategies that can minimize broadcasting operations while ensuring accurate position estimation.

\section{System Model}\label{sec:model}
In this section, we present the system model that is considered in our study. First, we theoretically model a \gls{cits} network as a time-varying Euclidean graph, whose nodes and edges represent vehicles and their communication links, respectively. Then, we define a performance metric that takes into account both the tracking errors and the vehicle positions. Finally, we describe the tracking system implemented by each node in the network and the communication channel through which vehicles' state information are broadcast.


\subsection{General Model}
\label{sub:general_model}

We represent a \gls{cits} network as a Euclidean graph $G=(V, E, r)$, i.e., an undirected graph whose vertices are points on a Euclidean plane~\cite{Balico:2018}. $V$ represents the set of nodes, $E$ represents the set of edges and $r$ is the node's communication range.
We say that two vehicles $v_i,v_j\in V$, $i\neq j$, are connected by the edge $<v_i,v_j>$ if the distance $d_{i,j}$ between them is shorter than the communication range $r$, i.e., $E=\{<v_i,v_j>:i\neq j, d_{i,j}<r\}$. 
Since the composition of the edge set depends on the positions of the vehicles, the topology of the network is time-varying, e.g., new edges can be activated or disabled according to how vehicles move. 

In our model, we assume that vehicles move in a two-dimensional space; while not always realistic, this hypothesis does not compromise the accuracy of our analysis.
To highlight the time dependency of the network, we denote by $G(t)=(V(t),E(t),r)$ the network graph at time $t$.
For simplicity, we assume that time is divided into discrete timeslots, so that the system evolves in steps.
Hence, we define the neighbor set $N_i(t)$ of $v_i$ during $t$ as the set of all the vehicles connected to $v_i$ by an edge in $E(t)$, i.e., $N_i(t)=\{v_j\in V(t):<v_i,v_j>\in E(t)\}$.

\begin{figure}[t]
	\centering
	\setlength{\belowcaptionskip}{-5pt}
	\includegraphics[height=1.8in]{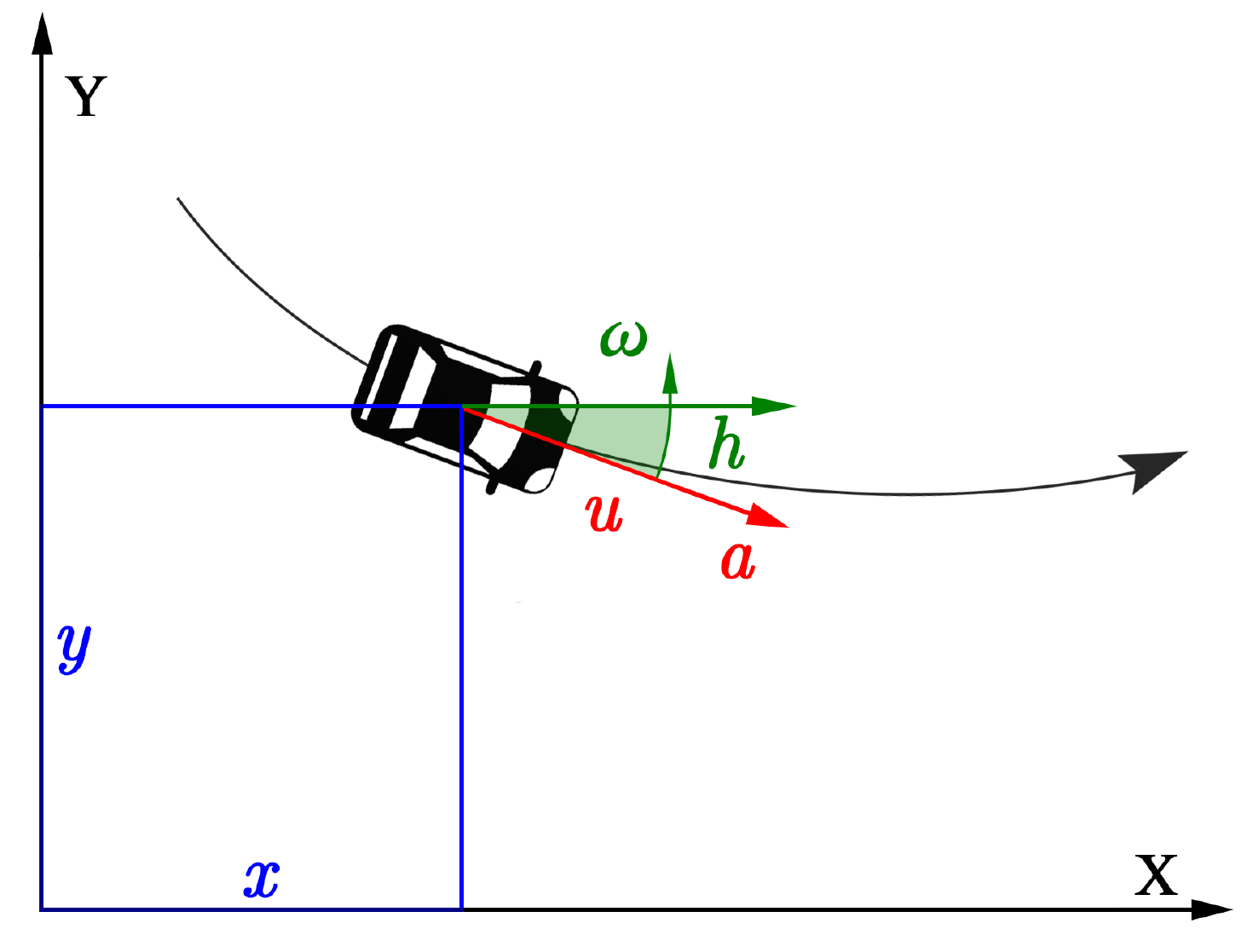}
	\caption{Graphical representation of the vehicle state $s(t)=\Big(x(t),y(t),h(t),u(t),a(t),\omega(t)\Big)$ at time $t$.}
	\label{fig:vehicle_state}
\end{figure}

The behavior of each vehicle $v_i$ in the network at time $t$ is represented by a 6-tuple $s(t)=\Big(x(t),y(t),h(t),u(t),a(t),\omega(t)\Big)$, which we call \emph{vehicle state}. In particular, $x$ and $y$ are the Cartesian coordinates of the vehicle on the road topology, $h$ is the vehicle's heading direction, $u$ and $a$ are the vehicle's tangent velocity and acceleration, respectively, and $\omega$ is the vehicle's angular velocity as exemplified in Fig.~\ref{fig:vehicle_state}. 
The physical distance between the positions of vehicles $v_i$ and $v_j$ at time $t$ is given by $d(s_1(t),s_2(t)) = \sqrt{(x_1(t) - x_2(t))^2 + (y_1(t) - y_2(t))^2}$. 

\subsection{Error Function}
\label{sub:error_function}

Consider a reference vehicle $v_i$ $\in$ $V(t)$, called the \emph{ego vehicle}, which tracks a group of other vehicles that we call \emph{target vehicles}. 
Hence, we denote by $\hat{N}_i(t)$ the subset of $N_i(t)$ containing the target vehicles, and by $\hat{s}_{i,j}(t)$ the state estimate of $v_j \in \hat{N}_i(t)$ performed by $v_i$ at time $t$. 
Under these hypotheses, the performance of the ego vehicle in terms of position estimation accuracy can be assessed by an \emph{error function} $\mathcal{F}(v_i,t)$, which is a weighted average of the position estimation errors made by the ego vehicle with respect to itself and all its target vehicles. Formally, we have
\begin{equation} \label{eq:error_function}
\begin{split}
\mathcal{F}(v_i,t) = \frac{1}{ |\hat{N}_i (t)|+1 } \left( \lambda_{i,i}(t) d(\hat{s}_{i,i} (t), s_i (t)) +
\sum_{ v_j \in \hat{N}_i (t) } \lambda_{i,j}(t) d(\hat{s}_{i,j} (t), s_j (t)) \right).
\end{split}
\end{equation}

In \eqref{eq:error_function}, $d(\hat{s}_{i,i} (t), s_i (t))$ and $d(\hat{s}_{i,j} (t), s_j (t))$ represent the error made by $v_i$ in estimating its own state $s_i(t)$ and the neighbor state $s_j(t)$, respectively, $|\hat{N}_i (t)|+1$ represents the total number of estimations carried out by $v_i$, and $\lambda_{i,k}(t)$ is the generalized logistic function defined as:
\begin{equation} \label{eq:sig_factor}
\lambda_{i,k}(t) = A_{\lambda} + \frac{E_{\lambda}-A_{\lambda}} {\left( C_{\lambda} + D_{\lambda} e^{ -B_{\lambda} (d(s_i(t), s_k (t))-d_0)} \right) ^{1 / \nu_{\lambda} }}.
\end{equation}
The parameters in \eqref{eq:sig_factor} characterize the function, also known as Richards' curve~\cite{richards1959flexible}, which was originally proposed for plant growth modeling. The values of all these parameters will be listed in Sec.~\ref{sec:analysis}. 

To evaluate the performance of the whole network, we define $\mathcal{F}(t)$ as the average of $\mathcal{F}(v_i,t)$ among all vehicles~$v_i \in V(t)$:

\begin{equation}\label{eq:err_function_vanet}
\mathcal{F}(t) = \frac{1}{|V(t)|}\sum_{v_i \in V(t)}\mathcal{F}(v_i,t).
\end{equation} 

\subsection{Tracking System}
\label{sub:tracking_system}

To minimize the positioning error defined in \eqref{eq:error_function}, the ego vehicle must estimate its state and the state of every other vehicle in the set $\hat{N}_{i}(t)$ in every timeslot. 
To reach this goal, the ego vehicle exploits both the information gathered by its on-board sensors and the information received from its neighbors through inter-vehicle communications. To allow the estimation of $s_i(t)$, we assume that, in every timeslot, the ego vehicle's on-board sensors provide a new observation $o(t)$ of $s_i(t)$. Hence, the ego vehicle can model the evolution of its own state through a Bayesian approach, obtaining the system

\begin{equation} \label{eq:bayesian_model}
\begin{cases}
s(t+1) = f(s(t)) + \zeta(t), \\
o(t) = m(s(t)) + \eta (t).
\end{cases}
\end{equation}
In \eqref{eq:bayesian_model}, the first equation describes the evolution of the vehicle state $s(t)$ over time, while the second equation describes the relation between $s(t)$ and the state observation $o(t)$. In particular, $f(\cdot)$ is a function describing the \gls{ctra} motion model given in \cite{Tsogas:2005}, while $m(\cdot)$ is a function representing the vehicle's measurement system. Moreover, $\zeta(t)$ and $\eta (t)$ represent the process and measurement noises, respectively, and are modeled as independent Gaussian processes with zero mean and covariance matrices $Q$ and $R$.
Once all the parameters in \eqref{eq:bayesian_model} are defined, the ego vehicle can estimate its own state by using a \gls{bf} algorithm. In our model, each vehicle implements a \gls{ukf} algorithm exploiting the \textit{sigma points} parameterization given in~\cite{Julier:2002}. 
By exploiting the \gls{ukf} and the system equations given in \eqref{eq:bayesian_model}, the ego vehicle obtains the estimate $\hat{s}_{i,i}(t)$ of its own state $s(t)$ and the related covariance matrix $P_{i,i}(t)$, which represents the uncertainty of the state estimation, in each timeslot $t$.

To allow $v_i$ to estimate the states of the other network nodes, each vehicle $v_j$ $\in$ $V(t)$ can transmit the estimate $\hat{s}_{j,j}(t)$ of their own state and the related uncertainty $\hat{P}_{i,i}$. The time frame by which new transmissions are initiated depends on the selected broadcasting strategy, as described in Sec.~\ref{sec:comm_strategy}.
Each message transmitted by $v_j$ is received by all the vehicles in $N_j(t)$ after a certain communication delay (provided that the transmission is not interfered, as we will  explain later).
Whenever the ego vehicle gets a message from another node that was not previously in its neighbor set, it initializes a new \gls{ukf} having as initial state and uncertainty the received state and covariance matrix, respectively. 
The new filter propagates the initial state over time by evolving the model blindly, and is then updated when new information is received.
If a vehicle $v_i$ does not receive state updates from a neighbor $v_j$ for a period longer than $\Delta_{track}$, it stops to track the considered target, i.e., $v_j$ is removed from the set $N_i$.

\subsection{Channel Access Model}
\label{sub:channel_model}

Inter-vehicle communications are modeled following the IEEE 802.11p standard, which defines the Physical (PHY) and \gls{mac} layer features of the \gls{dsrc} transmission protocol~\cite{li2010overview}. 
\gls{dsrc} defines seven different channels at the PHY layer, each containing $n_{sc,tot}=52$ sub-carriers \cite{Jiang:2008}. For simplicity of discussion, we assume that only a limited number of sub-carriers $n_{sc}\leq n_{sc,tot}$ can be used for broadcasting state information messages, while the rest is reserved for other applications.

\gls{dsrc} implements the \gls{csmaca} scheme at the \gls{mac} layer, where nodes listen to the wireless channel before sending. 
We consider an ideal 1-persistent \gls{csmaca} scheme, capable of successfully arbitrating the channel access among in-range vehicles in such a way that one single transmission per sub-carrier and timeslot is enabled, even in case of multiple potential transmitters. However, we assume that collisions can still occur among out-of-range vehicles that transmit towards the same  receiver, an issue known in the literature as \emph{hidden node} problem. Therefore, the transmission from a vehicle $v_a$ to a vehicle $v_b$ will suffer from a hidden terminal collision if any of $v_b$'s neighbors that are out of $v_a$'s range start a transmission that overlaps in time and frequency with $v_a$'s signal.
We also design and implement a congestion control algorithm to reduce the channel collision probability. More details will be given in Sec.~\ref{sec:cong_con}.

\section{Broadcasting Strategies} \label{sec:comm_strategy}

In this section, we describe the communication strategies that are used to regulate the inter-vehicular communications in our model. In particular, two different solutions are considered, namely \gls{pb}, which is already implemented by most \gls{cits} applications, and \gls{etb},  our original proposal.

\subsection{Benchmark: Periodic Broadcasting (PB)}\label{sub:time_strategy}

The \gls{pb} strategy represents the benchmark solution of our analysis. In the \gls{pb} scenario each vehicle chooses a constant \textit{inter-transmission period} $T_{period}$, so that its communication process follows an almost regular time-frame. 
Reducing $T_{period}$ would allow for a reduction of the \textit{misdetection} probability, which is the probability that a neighbor $v_j$ belongs to $N_i$ but not to $\hat{N}_i$, at the expense of increasing the probability of channel access collisions. The \textit{false detection} probability, i.e., the probability that a neighbor $v_j$ belongs to $\hat{N}_i$ but not to $N_i$, should follow the same trend.
In particular, a new transmission is allowed each time a \textit{new neighbor} is sensed and no transmissions were initiated in the previous two timeslots. 
The strategy is described by Alg.~\ref{alg:policy_0}.
\begin{algorithm}[t!]
	\footnotesize
	\setstretch{1.3}
	\caption{\footnotesize Periodic Broadcasting (PB) strategy}
	\label{alg:policy_0}
	\begin{algorithmic}[1]
		\REQUIRE $T_{last-tx} > 0 $, $\hat{s}_{i,i}(t) \in \mathbb{R}^6 $, $\hat{s}^p_{i,i}(t) \in \mathbb{R}^6 $, $\textit{new neighbor} \in \{\textit{True, False}\}$
		\ENSURE $ \textit{transmit} \in \{\textit{True, False}\}$
		
		\STATE $\textit{transmit} \gets \textit{False}$
		\STATE $T_{last-tx} \gets T_{last-tx} + T_t$	
		
		\IF{$T_{last-tx} > T_{period}$}
		\STATE $\textit{transmit} \gets \textit{True}$
		\ELSIF{\textit{new neighbor} \AND $T_{last-tx}> 2T_t$ }
		\STATE $\textit{transmit} \gets \textit{True}$
		\ENDIF
		
		\IF{\textit{transmit}}
		\STATE $T_{last-tx} \gets \max\{T_{last-tx}-T_{period}, 0\}$ 
		\STATE $ \hat{s}^p_{i,i}(t) \gets \hat{s}_{i,i}(t)$
		\ENDIF
		\RETURN \textit{transmit}
	\end{algorithmic}
\end{algorithm} 

\subsection{New Proposal: Error Threshold Broadcasting (ETB) }\label{sub:error_strategy}

In the \gls{etb} scenario each vehicle chooses an \textit{error threshold} $E_{thr}$ and regulates its communication behavior so that the overall position estimation error never exceeds $E_{thr}$.
To reach this goal, the \textit{ego vehicle} defines an additional \gls{ukf}, which replicates the \gls{ukf} operations of all the neighbor nodes that are tracking the \textit{ego vehicle} itself.
This filter propagates the \textit{ego vehicle}'s state by using only its \textit{predictive step} with no sensor input, as done by the other vehicles.
Each time the \textit{ego vehicle} triggers a new communication, the filter state is updated mimicking the operation performed by neighbor vehicles upon reception of the packet. Hence, at each timeslot $t$ the \textit{ego vehicle} knows both the \textit{a posteriori} state estimate $\hat{s}_{i,i}(t)$, which is the output of its main filter, and the \textit{a priori} state estimate $\hat{s}^p_{i,i}(t)$, which is the output of its purely predictive filter and represents the state estimate of $v_i$ made by its neighbor vehicles.
At each timeslot, the two different estimates are compared: if the difference $d(\hat{s}_{i,i}(t), \hat{s}^p_{i,i}(t))$ exceeds $E_{thr}$, a new transmission is initiated.
We observe that, as before, the communication process can vary according to some specific events. A maximum \textit{inter-transmission period} $T_{max}$ is defined to mitigate the undetection of new neighbors, and additional transmissions are initiated in case new neighbors are detected. This strategy is described by Alg.~\ref{alg:policy_1}.

\begin{figure}[t!]
	\centering
	\begin{subfigure}{.45\textwidth}
		\centering
		\includegraphics[height=1.8in, width=2.8in]{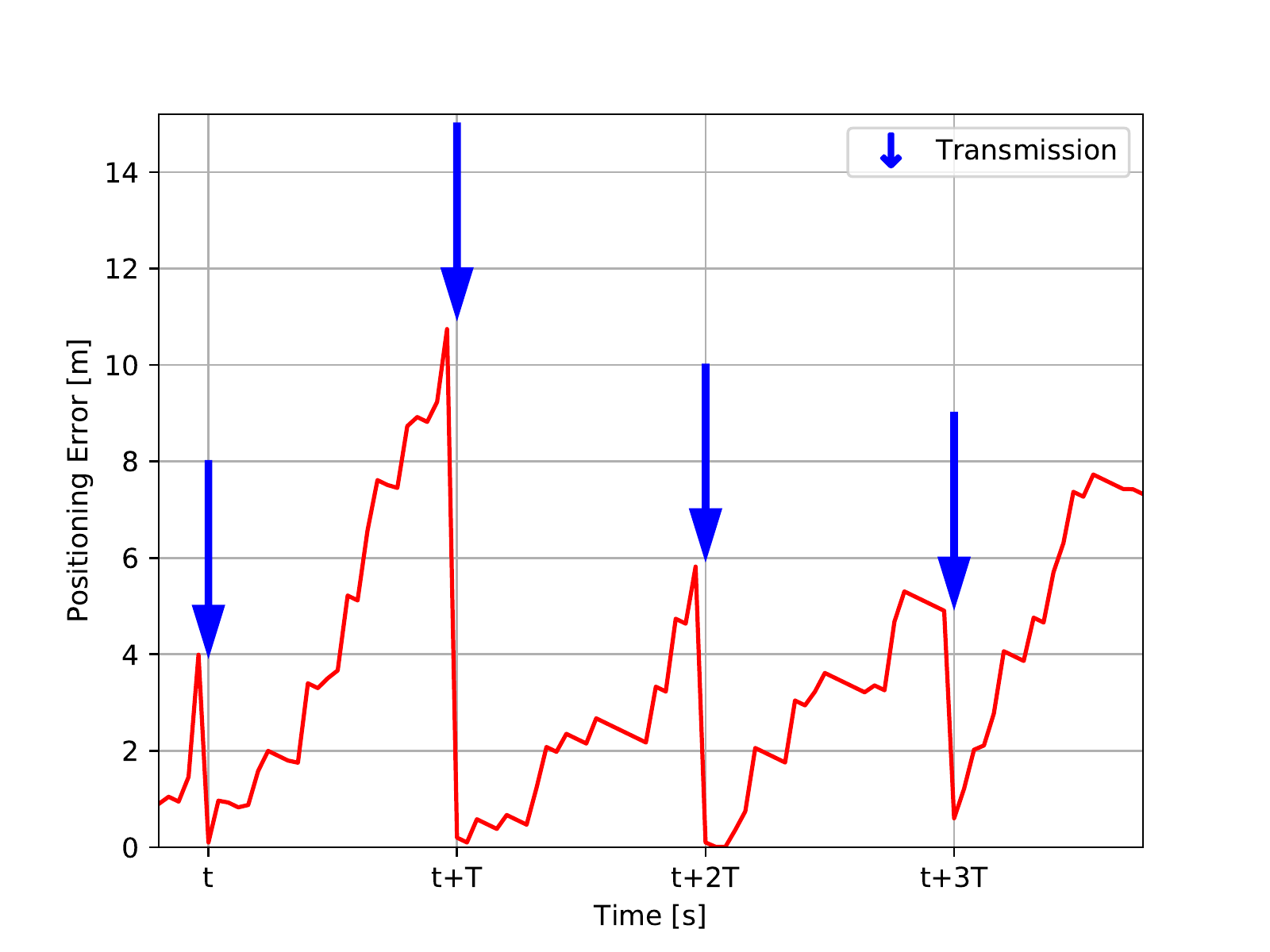}
		\caption{{PB strategy}.}
		\label{fig:policy0_error}
	\end{subfigure}
	\begin{subfigure}{.45\textwidth}
		\centering
		\includegraphics[height=1.8in, width=2.8in]{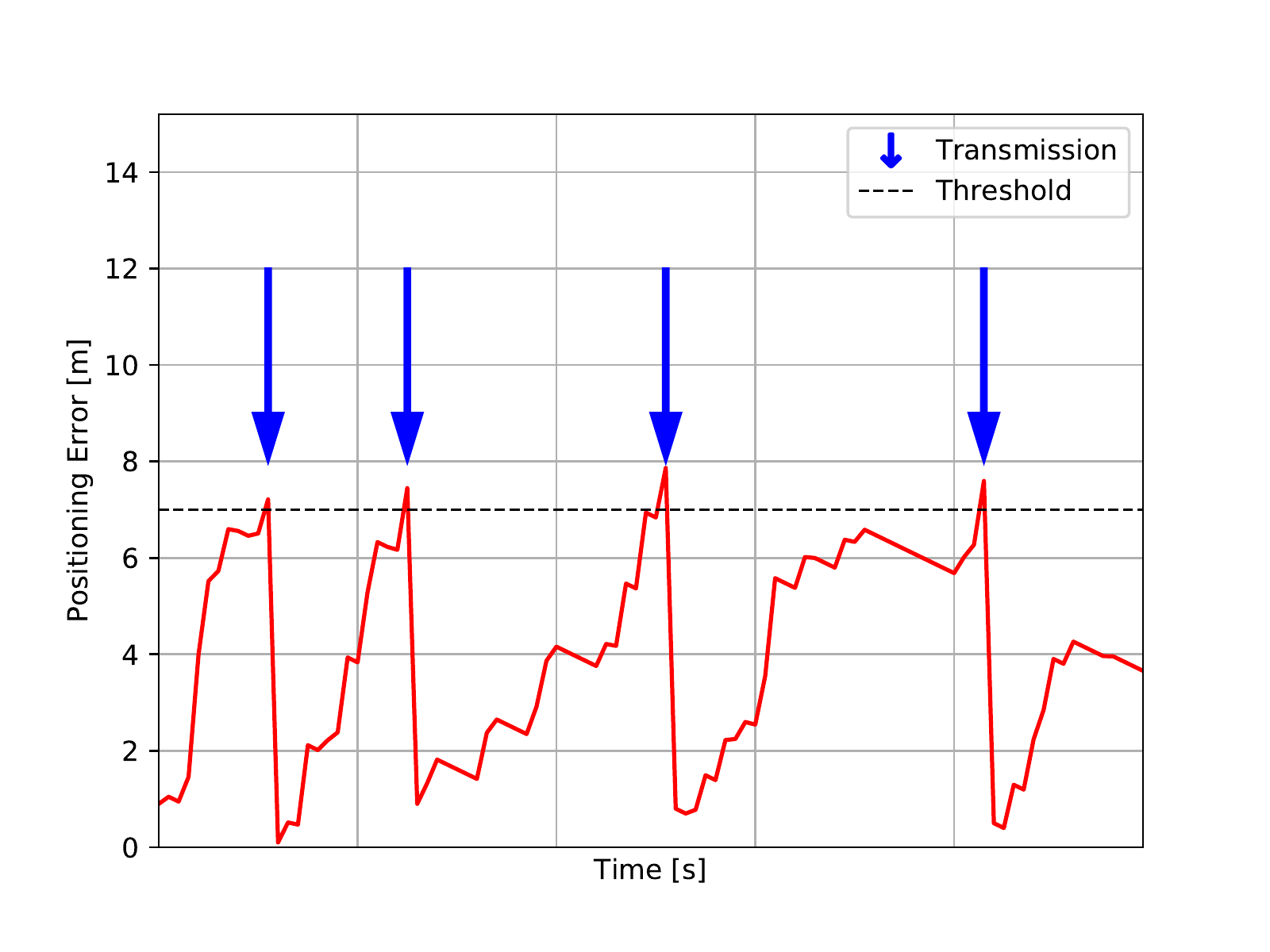}
		\caption{{ETB strategy}.}
		\label{fig:policy1_error}
	\end{subfigure}
\caption{Position Error evolution.}
\label{fig:error_evolution}
\end{figure}

An intuitive understanding of the \gls{pb} and \gls{etb} dynamics is provided by Fig.~\ref{fig:policy0_error} and Fig.~\ref{fig:policy1_error}, which represent the evolution of the position error $d(\hat{s}_{i,i}(t), \hat{s}_{i,i}^p(t))$ according to the transmission process in the two cases.
In the \gls{pb} scenario, we can observe that new transmissions are initiated in a regular fashion, regardless of the value of $d(\hat{s}_{i,i}(t), \hat{s}_{i,i}^p(t))$. Instead, in the \gls{etb} scenario, new transmissions are initiated only when $d(\hat{s}_{i,i}(t), \hat{s}_{i,i}^p(t))$ is above a certain threshold. 

\begin{algorithm}[t!]
\footnotesize
\setstretch{1.3}
	\caption{\footnotesize Error Threshold Broadcasting (ETB) strategy}
	\label{alg:policy_1}
	\begin{algorithmic}[1]
		\REQUIRE $T_{last-tx} > 0 $, $\hat{s}_{i,i}(t) \in \mathbb{R}^6 $, $\hat{s}^p_{i,i}(t) \in \mathbb{R}^6 $, $\textit{new neighbor} \in \{\textit{True, False}\}$
		\ENSURE $ \textit{transmit} \in \{\textit{True, False}\}$
		
		\STATE $\textit{transmit} \gets \textit{False}$
		\STATE $T_{last-tx} \gets T_{last-tx} + T_t$	
		\IF{$d(\hat{s}_{i,i}(t), \hat{s}^p_{i,i}(t)) > E_{thr}$ \OR $T_{last-tx} > T_{max}$ \OR (\textit{new neighbor} \AND $T_{last-tx}> 2T_t$)}
		\STATE $\textit{transmit} \gets \textit{True}$
		\ENDIF
		\IF{\textit{transmit}}
		\STATE $T_{last-tx} \gets \max\{T_{last-tx}-T_{period}, 0\}$ 
		\STATE $ \hat{s}^p_{i,i}(t) \gets \hat{s}_{i,i}(t)$
		\ENDIF
		\RETURN \textit{transmit}
	\end{algorithmic}
\end{algorithm}

In both the \gls{pb} and \gls{etb} scenarios, the best strategy setting would require to determine the optimal value of $T_{period}$ and $E_{thr}$, respectively. 
Such values can be computed through an exhaustive and computationally heavy approach that iterates on all the possible values of the number of available sub-carriers, the vehicular density, the characteristics of the road map, and other automotive-specific parameters, or with less resource-heavy congestion control techniques. In the next section, we will show how to adapt existing congestion control techniques to set both $T_{period}$ and $E_{thr}$ to minimize channel congestion.

\section{Congestion Control}\label{sec:cong_con}

Considering the dynamic nature of vehicular networks, the potential of the broadcasting strategies described in Sec.~\ref{sec:comm_strategy} can be fully expressed when coupled with congestion control  mechanisms that regulate information distribution as a function of the network load and minimize the packet collision probability.
In Sec.~\ref{sub:sensing} we present a benchmark congestion control mechanism, which we call \gls{cscc}.
The \gls{cscc} scheme is based on the  LIMERIC protocol~\cite{bansal:2013}, which will be reviewed in the following and, like most state-of-the-art approaches, relies on channel sensing.
We observe that channel sensing based mechanisms such as \gls{cscc} present several limitations, especially in highly dynamic scenarios.
To address these issues, in Sec.\ref{sub:adhoc} we design an alternative congestion control approach, which we call \gls{nacc}, exploiting the network topology information to reduce the packet collision probability. We remark that such congestion control algorithms can be used with either broadcast strategy, though \gls{nacc} has been designed for \gls{etb} and, hence, may underperform when combined with the \gls{pb} strategy. 

\subsection{Benchmark: Channel Sensing Congestion Control (CSCC)} \label{sub:sensing}

In the \gls{cscc} scenario, each vehicle constantly listens to the wireless channel and estimates the amount of resources that it is allowed to use to avoid congestion. We consider a vehicle $v \in V$, which is assigned to subcarrier $c_v \in \{0,1,..., n_{sc}-1 \}$. In each timeslot, $v$ senses the channel and determines if a new transmission has been initiated. The fraction of time during which the channel is sensed busy in the last $N_{cbr}^{avg}$ timeslots is called \gls{cbrl}. Every $N_{cbr}^{update}$ timeslots, the value of \gls{cbrl} is smoothed as
\begin{equation}
CBR_{vehicle} = 0.5 \cdot CBR_{vehicle} + 0.5 \cdot CBR_{local}.
\label{eq:cbr_v}
\end{equation}
The value of the \gls{cbrv} given by \eqref{eq:cbr_v} represents the channel occupancy sensed by $v$ over the subcarrier $c_v$. In the \gls{cscc} approach, each network node aims at keeping the value of \gls{cbrv} as close as possible to a target value, which is called \gls{cbrt}. Practically, every $N_{cbr}^{update}$ timeslots, $v$ evaluates the difference between \gls{cbrv} and \gls{cbrt} and updates accordingly the values of $\rho$, which is the fraction of time that $v$ can exploit to transmit over the wireless channel. This procedure is reported in Algorithm \ref{alg:rho}. 
\begin{algorithm}[t!]
\footnotesize
\setstretch{1.3}
	\caption{\footnotesize LIMERIC protocol~\cite{bansal:2013}}
	\label{alg:rho}
	\begin{algorithmic}
		\REQUIRE $CBR_{vehicle} > 0$, $\rho >0 $
		\ENSURE $ \rho>0$
		
		\IF{$CBR_{target} - CBR_{vehicle} > 0$}
		\STATE $\delta = \min\left(\beta \cdot (CBR_{target} - CBR_{vehicle}), \delta_{max}\right)$
		\ELSE
		\STATE $\delta = \max\left(\beta \cdot (CBR_{target} - CBR_{vehicle}), \delta_{min}\right)$
		\ENDIF
		\STATE $\rho = \left[(1-\alpha) \cdot \rho + \delta\right]_{\rho_{min}}^{\rho_{max}}$
		\RETURN $\rho$
	\end{algorithmic}
\end{algorithm}

When $v$ adopts the \gls{pb} strategy, the value of $T_{period}$ is updated as
$T_{period} = \frac{1}{\rho}$. In case $v$ is adopting the \gls{etb} strategy instead, the value of $\rho$ should be associated to a specific \textit{error threshold} $E_{thr}$, as explained in the last part of the current section.  We also limit the possible values of $\rho$ to the interval between $\rho_{min}$ and $\rho_{max}$, as $[x]_a^b=\min(\max(x,a),b)$.

\subsection{New Proposal: Neighbor Aware Congestion Control (NACC)} \label{sub:adhoc}

In the \gls{nacc} approach, each vehicle computes the value of $\rho$ as a function of its knowledge about neighbors' positions. In particular, vehicles can increase or decrease the channel occupancy with the aim of minimizing the packet collision probability. We start by theoretically modeling the communication that takes place in a group of vehicles when \gls{csmaca} is implemented at the \gls{mac} layer. Then, we describe how a user can estimate the number of neighbors that may potentially result in packet collisions. Finally, we find a relation between the vehicular density sensed by a user and the packet collision probability itself. 

\subsubsection{CSMA/CA Analysis}

We saw in Sec.~\ref{sub:channel_model} that vehicles access the channel following a 1-persistent \gls{csmaca} protocol. We now consider a population of $N$ vehicles that share the same subcarrier $c_v \in \{0,1,..., n_{sc}-1 \}$, which is supposed to be reserved to this set of vehicles. 
First, we assume that all the vehicles are always mutually in-range, i.e., at a distance lower than $r$.
Then, we assume that all the vehicles that are not already trying to transmit attempt to access the channel with the same transmission probability $\rho$. If $x_{t-1}$ vehicles are already trying to access the system at timeslot $t$, the probability that the number of new arrivals $a_t$ is equal to $a$ is given by: 
\begin{equation} 
P(a_t=a|x_{t-1} = i, \rho, N) = \begin{cases} 
\binom{N-i}{a} \rho ^a (1 - \rho)^{N-i - a}, & 0 \leq a \leq N-i; \\ 
0, & a>N-i. 
\end{cases} 
\end{equation} 
The number of vehicles in the access list after the timeslot is then $x_t=\max(x_{t-1}+a_t-1,0)$, as $a_t$ new vehicles are now attempting a transmission and at most one managed to transmit. Since the probability of accessing the channel is the same for all vehicles, the probability that a specific vehicle trying to access the channel will transmit is $(x_t+1)^{-1}$. 

We observe that, if $\rho$ and $N$ are fixed, the channel dynamics at the end of any timeslot $t$ are completely characterized by the number of users that need to transmit, i.e., the queue size $x_t$. Hence, we can describe the overall system by a \textit{Markov Chain}, whose states $x_t$ are in the set $X=\{0,\ldots,N-1\}$ and whose transition probability matrix $\mathbf{T}(\rho, N)$ is given by:
\begin{equation}
T_{i,j}(\rho, N) = \begin{cases}
0, & j < i-1; \\
P(a_t = j-i+1|x_{t-1} = i, \rho, N), & 0<i<N, \: i-1 \leq j < N; \\
P(a_t = j+1|x_{t-1} = i, \rho, N), & i = 0, \: 0<j<N; \\
P(a_t \leq 1|x_{t-1} = i, \rho, N), & i = 0, \: j = 0.

\end{cases}
\end{equation}
Given $\mathbf{T}(\rho, N)$, we can compute the steady state vector $\overline{\Pi}(\rho, N)=[ \Pi_0(\rho, N), \Pi_1(\rho, N), ..., \Pi_{N-1}(\rho, N) ] $. Then, we can compute the probability of different transmission events. In particular the probability that during a generic timeslot $t$ no transmissions are initiated is given by

\begin{equation}
P(x_{t-1} = 0, a_t = 0 | \rho, N) = \Pi_0 (\rho, N) \cdot P(a_t = 0|x_{t-1}=0, \rho, N).
\label{eq:no_tx}
\end{equation}

\subsubsection{Vehicle Position Distribution}

We recall that our objective is to minimize the number of packet collisions, which in our model are caused only by the \textit{hidden terminal} problem. 
To compute the collision probability in the described scenario, we should estimate how many  neighbors of the target receiver $v_b$ can interfere. We denote this value by $N_{ht}$. Assuming that the vehicular density in the communication area of $v_b$ is constant, we can estimate $N_{ht}$ as
\begin{equation}
\hat{N}_{ht} = \frac{N_b+1}{n_{sc}} \frac{E\left[ \mathcal{A}(d) \right]}{\pi r^2} .
\label{eq:hidden_vehicles}
\end{equation}

\begin{figure}[t!]
	\centering
	\includegraphics[height=2.8in,  width=2.8in]{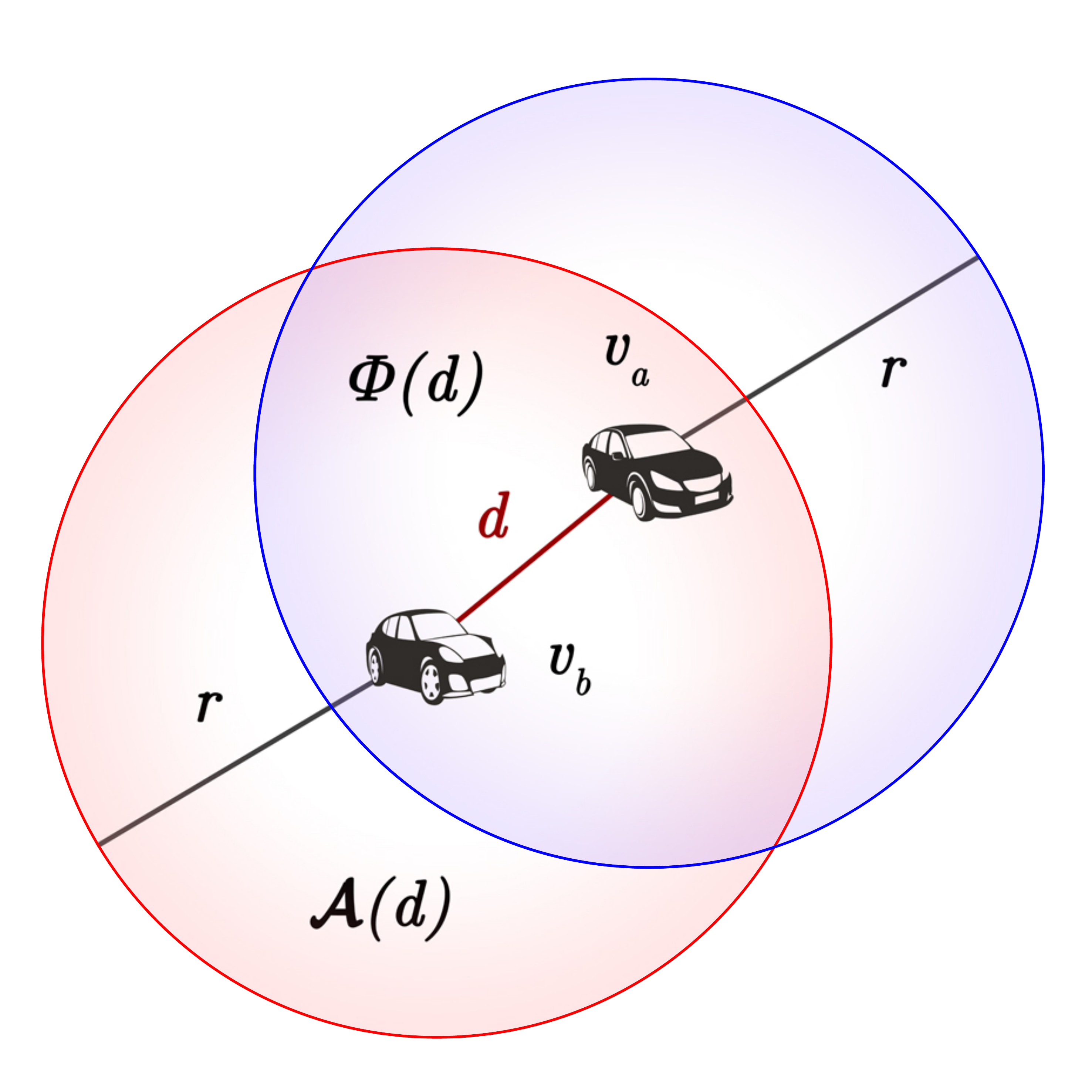}
	\caption{Intersection $\Phi(d)$ of the communication ranges of $v_a$ and $v_b$}
	\label{fig:area}
\end{figure}

In \eqref{eq:hidden_vehicles},$\frac{N_b+1}{n_{sc}}$ is the estimate of the number of vehicles contained in the communication area of $v_b$ that are using the same subcarriers of $v_a$,  $d$ is the distance between $v_a$ and $v_b$ while $ \mathcal{A}(d) $ is the area within the coverage of $v_b$ but not of $v_a$. In other words, $\mathcal{A}(d)$ is the area from which a transmission would be hidden from $v_a$, possibly causing a hidden node collision. 
Let us define by  $\Phi(d)$  the intersection of the communication areas of $v_a$ and $v_b$, so that $\Phi(d) = \pi r^2 - \mathcal{A}(d)$.  
A graphical representation of $\Phi (d)$ is reported in Fig.~\ref{fig:area} while its mathematical expression is given~by

\begin{equation}
\Phi (d) = 2 r \left( r \arccos\left(\frac{d}{2 r}\right) - \frac{d}{2}  \sqrt{1 - \left(\frac{d}{2 r}\right)^2}  \right).
\label{eq:area_inter}
\end{equation}

Assuming that the geographical distribution of the nodes can space can be modeled as as a \gls{ppp}, the probability distribution of $d$ is equal to
\begin{equation}
f_d(d) = \frac{2 d}{r^2}.
\label{eq:d_distr}
\end{equation}
Given \eqref{eq:area_inter} and \eqref{eq:d_distr}, the mean value of $\Phi(d)$ can be computed as 
\begin{equation}
E\left[ \Phi(d) \right]  = \int \Phi (d) f_d (d) dd  = r ^ 2 \left( \pi - \frac{3 \sqrt{3}}{4} \right).
\end{equation}
Recalling that $\Phi(d) = \pi r^2 - \mathcal{A}(d)$, we can write 
\begin{equation}
E\left[ \Phi(d) \right]  = r ^ 2 \left( \pi - \frac{3 \sqrt{3}}{4} \right) = \pi r^2 - E\left[ \mathcal{A}(d) \right],
\end{equation}
so that $E\left[ \mathcal{A}(d) \right]=\frac{3 \sqrt{3}}{4}r^2$. Replacing $E\left[ \mathcal{A}(d) \right]$ in  \eqref{eq:hidden_vehicles} we finally obtain the expression
\begin{equation}
\hat{N}_{ht} = \frac{N_b+1}{n_{sc}} \frac{3 \sqrt{3}}{4 \pi}.
\end{equation}

\subsubsection{Packet Collision Probability}

We consider a vehicle $v$ that is tracking $\hat{N}$ neighbors (the value of $\hat{N}$ depends on the overall system dynamics and may differ from $N$, which instead represents the true number of neighbors that are in the communication area of $v$).
Suppose that $v$ starts a new transmission during a generic timeslot $t$. On average there are $\hat{N}_{ht} = \frac{\hat{N}+1 }{n_{sc}} \left(1-\frac{\overline{\Phi}}{\pi r^2} \right)$ vehicles which can interfere with the communication.
Hence, according to our channel model, the probability that the transmission will not fail corresponds  to the probability that none of those $\hat{N}_{ht}$ interfering nodes transmits during $t$.
If we assume that the considered $\hat{N}_{ht}$ vehicles have the same transmission probability $\rho$ and do not interact with other network nodes during $t$, the packet collision probability $P_{coll}$ can be derived from \eqref{eq:no_tx}, obtaining
\begin{align}
P_{coll} (\rho, \hat{N}_{ht})&=1-\Pi_0(\rho,\hat{N}_{ht})(1-\rho)^{\hat{N}_{ht}}.
\label{eq:coll_prob}
\end{align}
To reduce the number of collisions, the vehicle $v$ with $N$ neighbors should have a transmission probability $\rho$ such that $P_{coll}$ equals a predetermined threshold $P_{thr}$. In other words, $v$ chooses $\rho$ so that the difference between $P_{coll}$ and $P_{thr}$ is minimized, which means
\begin{equation}
\rho = \argmin_{\rho} \left(\left| P_{coll}(\rho, \hat{N}_{ht}) - P_{thr} \right|\right). 
\label{eq:ad_hoc}
\end{equation}

The above procedure is described in Alg.~\ref{alg:nacc}. Following the \gls{nacc} protocol, each vehicle $v$ changes the value of $\rho$ according to the vehicular density in its surroundings. In particular, in case the vehicle is using the \gls{pb} strategy, the value of $T_{period}$ is updated as $T_{period} = \frac{1}{\rho}$.
We highlight that, by adjusting the value of $T_{period}$ in this way, we violate the assumption regarding the distribution of the packet inter-transmission time considered in the definition of the system Markov model. Indeed, with the \gls{pb} strategy, the time between two subsequent transmissions is constant rather than geometrically distributed, while in the \gls{etb} strategy scenario it depends on the position error evolution. This approximation may impair the performance of our congestion control mechanism. In particular, we expect to observe a significant performance reduction in the case of the \gls{pb} strategy.

\begin{algorithm}[t!]
	\footnotesize
	\setstretch{1.3}
	\caption{\footnotesize NACC protocol}
	\label{alg:nacc}
	\begin{algorithmic}
		\REQUIRE $n_{sc} > 0 $, $\overline{\Phi} \geq 0$, $\hat{N} \geq 0$
		\ENSURE $ \rho>0$
		
		\STATE $\hat{N}_{ht} = \frac{\hat{N}+1}{n_{sc}} \left(1-\frac{\overline{\Phi}}{\pi r^2} \right)$
		
		\STATE $P_{coll} (\rho, \hat{N}_{ht}) = 1-\Pi_0 (\rho, \hat{N}_{ht}) P(a_t = 0 | x_{t-1} = 0, \rho, \hat{N}_{ht})$
		
		\STATE $\rho = \argmin \left(\left| P_{coll}(\rho, \hat{N}_{ht}) - P_{thr} \right|\right)$
		
		\RETURN $\rho$
	\end{algorithmic}
\end{algorithm}

\subsection{Implementing Congestion Control for the \gls{etb} Strategy} \label{sub:adaptation}
Both the \gls{cscc} and the \gls{nacc} approaches improve the efficiency of the broadcasting strategies described in Sec.~\ref{sec:comm_strategy} by  adapting the inter-transmission period to the deployment scenario. 
As stated previously, to combine a congestion control scheme with the \gls{etb} strategy, we have to relate the inter-transmission period to the error threshold.
Practically, we need to build a map $\mathcal{F}$ such that the transmission period $T_{period} = \mathcal{F}(E_{thr})$ yields an average map estimation error close to $E_{thr}$. Unfortunately, the relation between $E_{thr}$ and $T_{period}$ is subject to multiple factors and cannot be easily modeled.
$\mathcal{F}$ depends on how the position estimation error of vehicles evolves in time, i.e., on both the road map and the users' behaviors.

To reach our goal, we hence resorted to a pragmatic approach. 
By simulating a purely predictive UKF in the considered scenario, we derive an empirical estimate of the statistical distribution $P(e_h\leq E_{thr})$ of the position estimation error $e_h$ after $h$ timeslots since the last update, for any $h \geq 0$.
Denoting by $\mathcal{H}$ the number of timeslots at which the error $e_h$ exceeds the threshold $E_{thr}$, we can set $T_{period}=E[\mathcal{H}] T_{t}$, where $T_t$ is the timeslot duration. 
Now, pretending the $e_h$ are independent random variables, the complementary cumulative distribution function of $\mathcal{H}$ can be expressed as 
\begin{equation}
P(\mathcal{H}>H) = \prod_{h=1}^H P(e_h \leq E_{thr}) 
\end{equation}
from which we easily get
\begin{equation}
T_{period} = T_t \sum_{H=1}^\infty \prod_{h=1}^H P(e_h \leq E_{thr}) 
\label{eq:errper_map}
\end{equation}

Equation \eqref{eq:errper_map} hence provides the desired map $\mathcal{F}$ from the error threshold $E_{thr}$ to the inter-transmission period $T_{period}$.
Such a function can also be used to determine the value $\rho$ of the broadcast policy ETB, which can be computed as follows:
\begin{equation}
\rho = \frac{1}{\mathcal{F}(E_{thr})}. 
\end{equation}
We highlight this approach requires that vehicles know the distribution of the position estimation error in the map. In a realistic scenario,  such information can be  provided to vehicles by the road infrastructure, or pre-programmed into the channel access algorithm (possibly with multiple choices, depending on the road conditions). The investigation of such aspects, however, is left to future work.  

\section{Performance Analysis and Simulation Results}\label{sec:analysis}

In this section we evaluate the performance of the proposed \gls{etb} strategy for broadcasting operations, compared to a traditional \gls{pb} approach. 
Moreover, we exemplify how the proposed \gls{nacc} mechanism  can improve the performance of the broadcasting strategies by exploiting network topology information, with respect to the benchmark \gls{cscc} scheme that relies only on channel sensing.
The results of our work are derived through a Monte Carlo approach, where $N_{\rm sim}$ independent simulations of duration $T_{\rm sim}=100$ s are repeated to get different statistical quantities of interest. The simulation parameters listed in Tab.~\ref{tab:general} are based on a realistic urban \gls{cits} scenario.

\paragraph{General parameters}
\label{par:general_parameters}
We use conservative IEEE 802.11p PHY and MAC layer parameter settings, which yield a maximum discoverable range of $r=140$ m~\cite{benin2012vehicular}, while the communication delay is set to $T_d =100$ ms, corresponding to one timeslot $T_t$.
When not implementing a congestion control scheme, the settings of both the \gls{pb}  and \gls{etb} strategies must be defined \textit{a priori}.
In our simulations we adopt an exhaustive approach and consider $N_{\rm set}=30$ different settings.
In particular, we make the inter-transmission period $T_{period}$ vary from 0 to 10 seconds while the error threshold $E_{thr}$ ranges between 0 and 42 meters.
Each choice involves a different trade-off between estimation accuracy and broadcasting overhead. 

\begin{figure}[b!]
	\centering
	\begin{subfigure}[t]{0.44\columnwidth}
		\centering
		\includegraphics[height=3in,  width=3in]{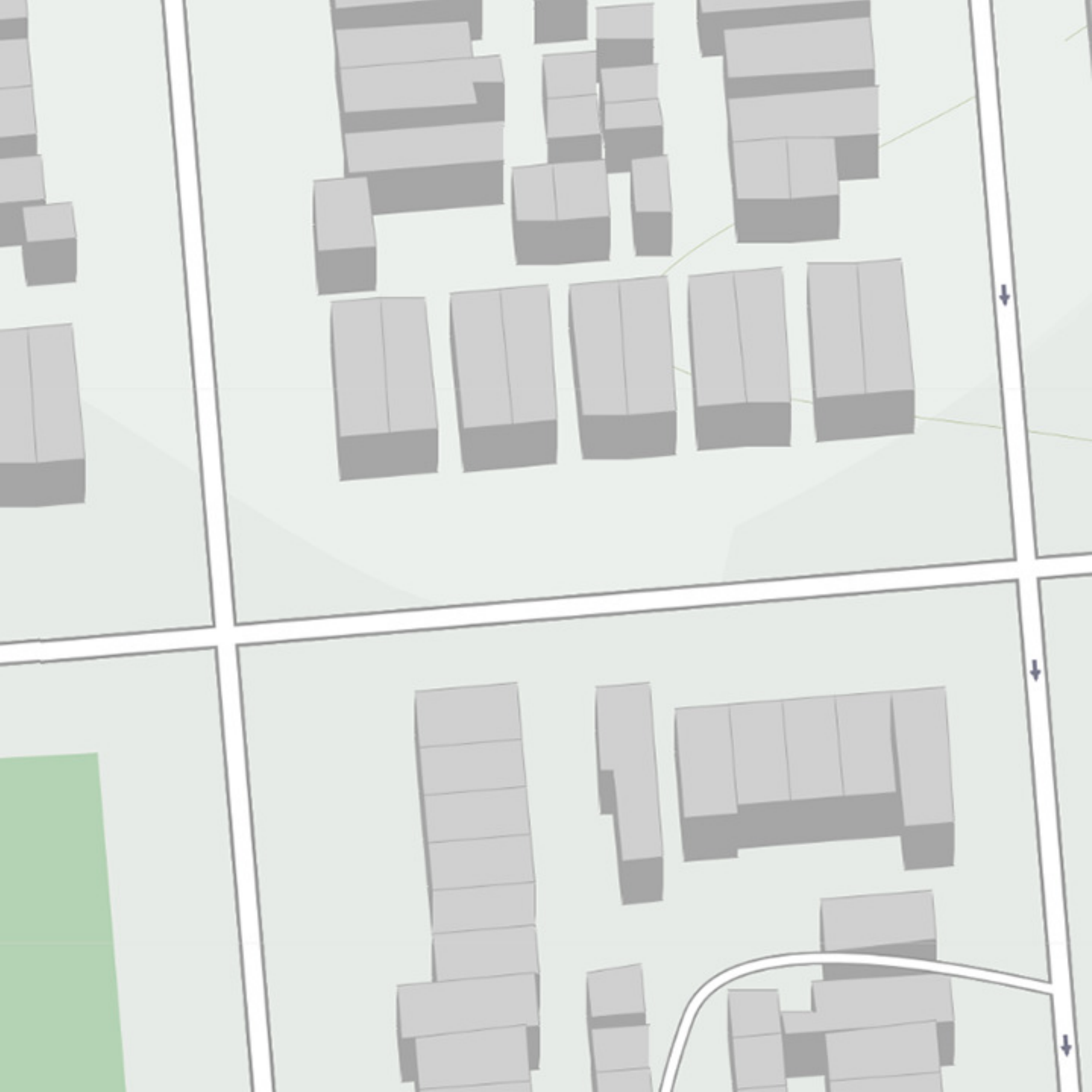}
		\caption{Openstreetmap scenario.}
		\label{fig:osm}
	\end{subfigure} 
	\begin{subfigure}[t]{0.44\columnwidth}
		\centering
		\includegraphics[height=3in,  width=3in]{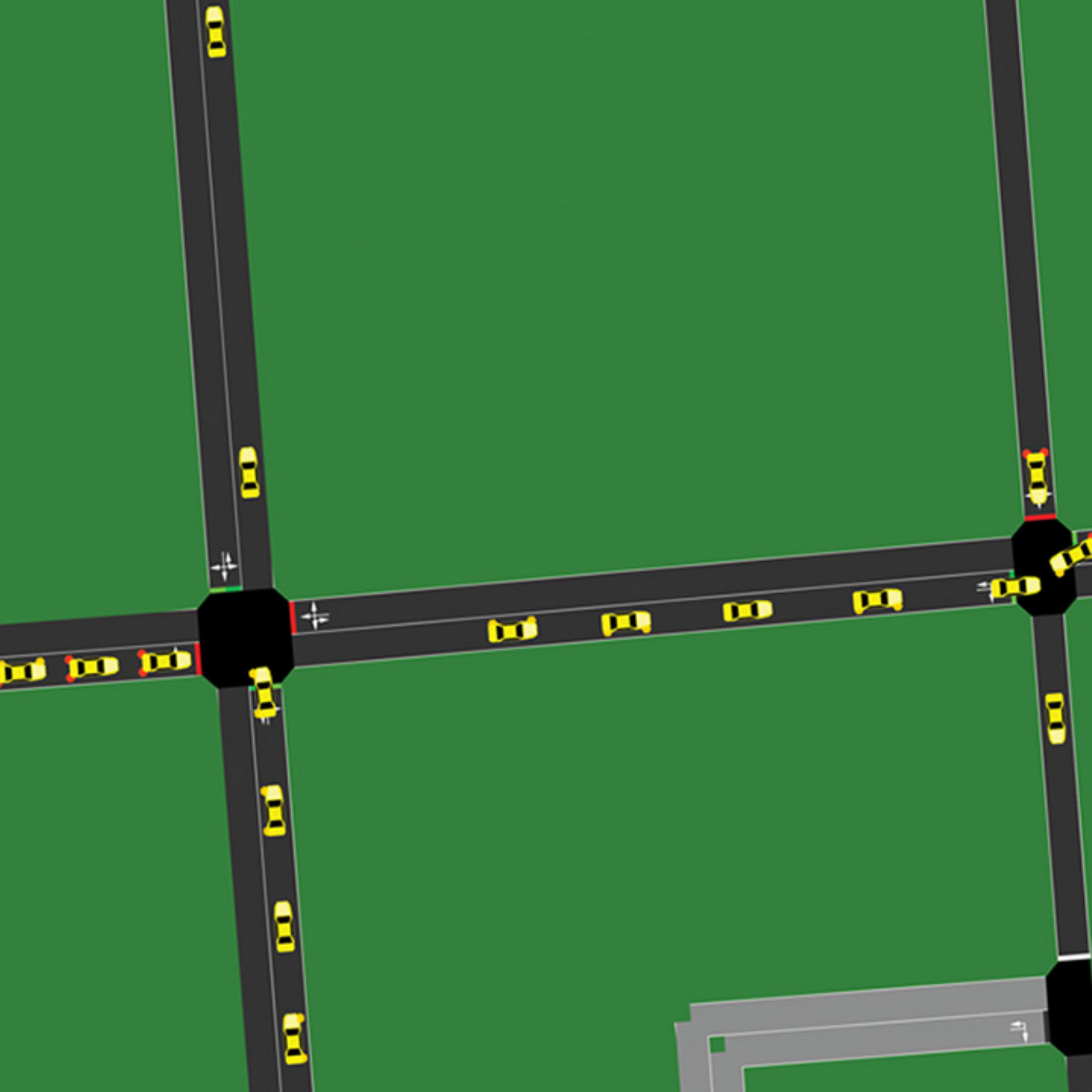}
		\caption{SUMO scenario.}
		\label{fig:sumo}
	\end{subfigure}
	\caption{Representation of a portion of the urban map considered for the performance evaluation.}
	\label{fig:map}
\end{figure}

For our simulations, we use real road map data imported from {OpenStreetMap} (OSM), an open-source software that combines wiki-like user generated data with publicly available information.
In particular, we consider the OSM map of New York City, as represented in Fig.~\ref{fig:osm}, so that to characterize a dynamic urban environment.
In order to consider realistic mobility routes that are representative of the behavior of vehicles in the network, we simulate the mobility of cars using SUMO, as represented in Fig.~\ref{fig:sumo}.
The vehicles move through the street network according to a \texttt{randomTrip} mobility model, which generates trips with random origins and destinations, and speeds that depend on the realistic interaction of the vehicle with the road and network elements.
The maximum speed is set to $v_{max}=13.89$ m/s, which is consistent with current speed limits.
Given $v_{max}$, we set $d_0=42$ m, which corresponds to the distance traveled in $3$ seconds by a vehicle running at the maximum speed. In this way, $d_0$ represents the maximum \emph{safety distance} that should be held in an urban scenario.
Following the work of \cite{Boban:2016}, we consider a vehicular density of $d_v=120$ vehicles/km$^2$ for medium traffic conditions.
Given the total road map area of A=$0.5168$ km$^2$, the number of vehicles deployed in the considered scenario is $|V|=62$.

As we assessed in Sec.~\ref{sec:model}, the behavior of each node can be fully represented by its state $s(t)$. Measurements of the components of $s(t)$ are affected by a non-negligible noise which is modeled as a Gaussian process with zero mean and covariance matrix $R$. The diagonal elements of $R$ are given in Tab.~\ref{tab:general} and are derived from the models in \cite{Kim:2014,Driver:2007,Falco:2017}. We define $Q = qI$, where $q$ is the process noise covariance parameter and $I$ denotes the identity matrix. 

Tab.~\ref{tab:general} also reports the parameters of the congestion control schemes from Sec.~\ref{sec:cong_con}. For what concerns the \gls{cscc} approach, we use the same parameters suggested in~\cite{bansal:2013}. Therefore, we set \gls{cbrt} to $0.68$, so that vehicles aim at occupying the channel about 68\% of the time, $\alpha = 0.1$, which ensures a sufficiently high convergence speed, and $\beta = (2-\alpha) /K$, so that the algorithm convergence is guaranteed for any $K$. We observe that $K$ represents the maximum number of users sharing the same communication channel that in our scenario is on average $|V|/n_{sc}$. For what concerns our proposed \gls{nacc} approach, we set the collision probability threshold to $P_{coll}=0.3$. We choose this value to allow a fair comparison between the \gls{cscc} and the \gls{nacc} approaches. Indeed, setting $\text{\gls{cbrt}}=0.68$ and $P_{coll}=0.3$, we obtain similar values of mean inter-transmission time $\overline{T}_{tx}$ for each combination of broadcasting strategy and congestion control mechanism. 

\begin{table}[t!]\vspace{12pt}
	\centering
	\setstretch{1.5}
	\tiny
		\caption{General parameters.}
	\label{tab:general}
	\begin{tabular}{@{}lll|lll@{}}
		\toprule
		Parameter & Value & Description & Parameter & Value & Description\\ \midrule
		$T_{\rm sim}$ & $100$ s & Simulation duration & $v_{max}$ & $13.89$ m/s & Maximum speed \\
		$N_{\rm sim}$ & $20$ & Number of runs & $d_0$ & $42$ m & Safety distance \\
		$T_t$ & $100$ ms & Timeslot duration & $A_S$ & $0.5168$ km$^2$ & Area size \\
		$T_{d}$ & $100$ ms & Communication delay & $d_v$ & $120$ vehicles/km$^2$ & Vehicular density \\
		$r$ & $140$ m & Communication range & $|V|$ & $62$ & Number of nodes \\	
		$n_{sc}$ & \{$2$, $4$, $6$, $8$, $10$\} & Number of subcarriers & $\rho_{max}$ & $1$ & Upper bound of $\rho$ \\	
		$n_{sc,tot}$ & $52$ & Maximum number of subcarriers & $\Delta_{track}$ & $10$ s & Maximum tracking duration \\	
		$E_{thr}$ & $\{0,\dots 42\}$ m & Error threshold & $\rho_{min}$ & $0.0006$ & Lower bound of $\rho$ \\
		$T_{period}$ & $\{0,\dots 10\}$ s & Inter-transmission period & $\delta_{max}$ & $1$ & Upper bound of $\delta$ \\
		$q$ & $1$ & Process noise parameter & $\delta_{min}$ & $-1$ & Lower bound of $\delta$ \\
		$R_{1,1}$ & $1.18535$ m$^2$ & Position accuracy along x & $K$ & $ |V| / n_{sc} $ & Maximum number of users \\
		$R_{2,2}$ & $1.18535$ m$^2$ & Position accuracy along y & $\alpha$ & $0.1$ & Algorithm speed parameter\\
		$R_{3,3}$ & $0.5$ (m/s)$^2$ & Speed accuracy & $\beta$ & $ (2-\alpha) /K$ & Algorithm convergence parameter \\
		$R_{4,4}$ & $0.39$ (m/s$^2$)$^2$ & Acceleration accuracy & $CBR_{target}$ & 0.68 & Target Channel Busy Ratio \\
		$R_{5,5}$ & $0.09211$ rad$^2$ & Heading accuracy & $P_{thr}$ & $0.3$ & Collision probability threshold \\
		$R_{6,6}$ & $0.01587$ (rad/s) $^2$ & Turn rate accuracy & \{$A_{\lambda}$, $B_{\lambda}$, $C_{\lambda}$, $D_{\lambda}$, $E_{\lambda}$, $\nu$\} & \{$1$, $0.05$, $1$, $1$, $0$, $0.2$\} & Logistic function parameters \\ 
		
		\bottomrule
	\end{tabular}
\end{table}

To evaluate the performance of the proposed broadcasting strategies in the simulations, we take into account four main factors, namely:
\begin{itemize}
	\item \emph{Average positioning error}, i.e., the average error of the ego vehicle when estimating its own position and that of its neighbors, which is given by \eqref{eq:error_function};
	\item \emph{95th percentile of the positioning error}, i.e., the positioning error threshold exceeded only by the worst 5\% of the vehicles;
	\item \emph{Detection error}, i.e., the sum of the misdetection (i.e., unknown vehicles in the ego vehicle communication area) and false detection (i.e., vehicles that are believed to be in the neighborhood but are actually beyond the communication range) event probabilities;
	\item \emph{Packet collision rate}, i.e., the average number of packet collisions per vehicle and per second that occur because of the hidden terminal problem. \\
\end{itemize}

\subsection{Theoretical Analysis and Validation}
\label{sub:validation}

\begin{figure}[h]
	\centering
	\includegraphics[width=4in]{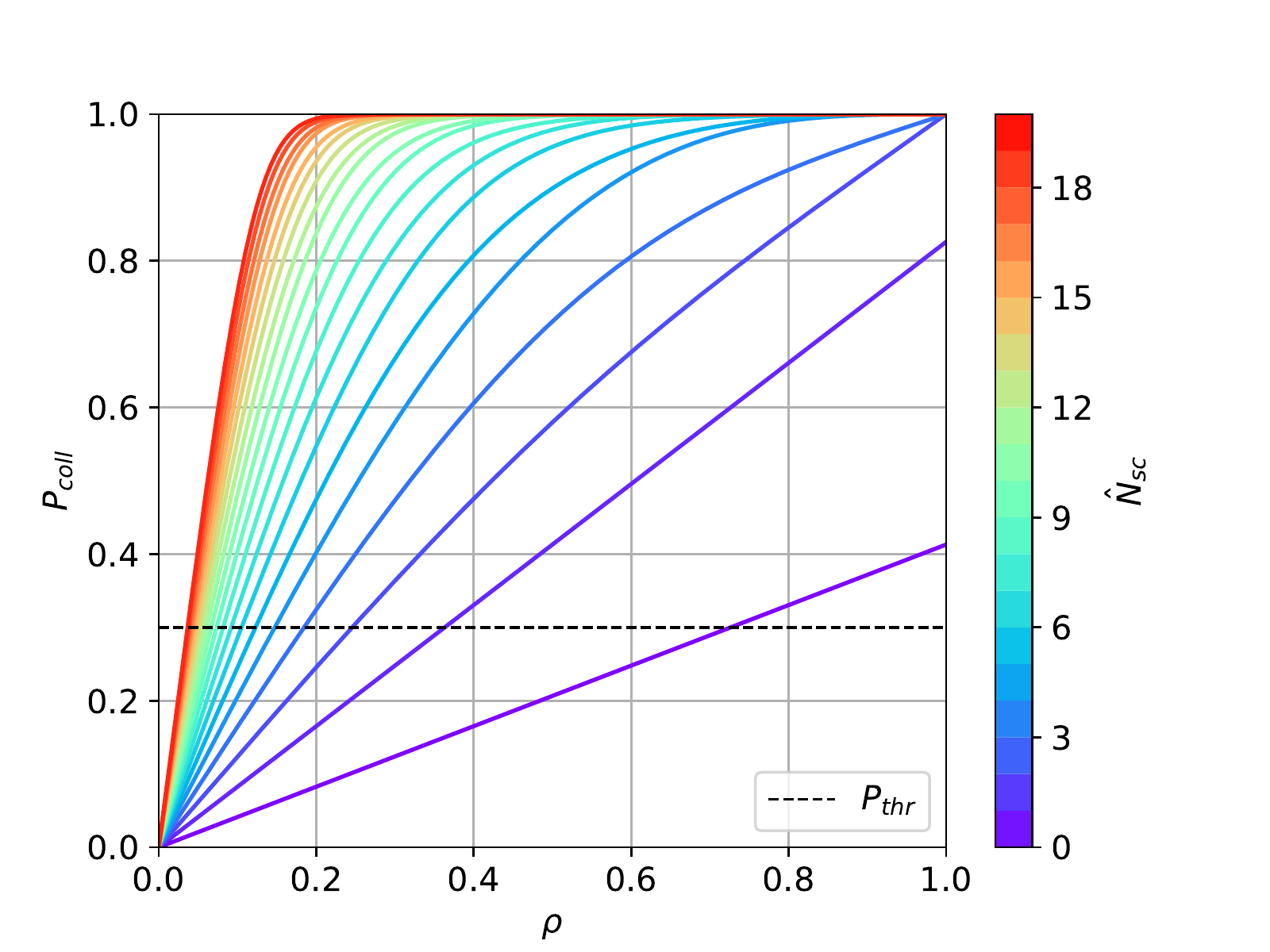}
	\caption{Collision probability.}
	\label{fig:collision_prob}
\end{figure} 
\begin{figure}[h]
	\centering
	\includegraphics[width=4in]{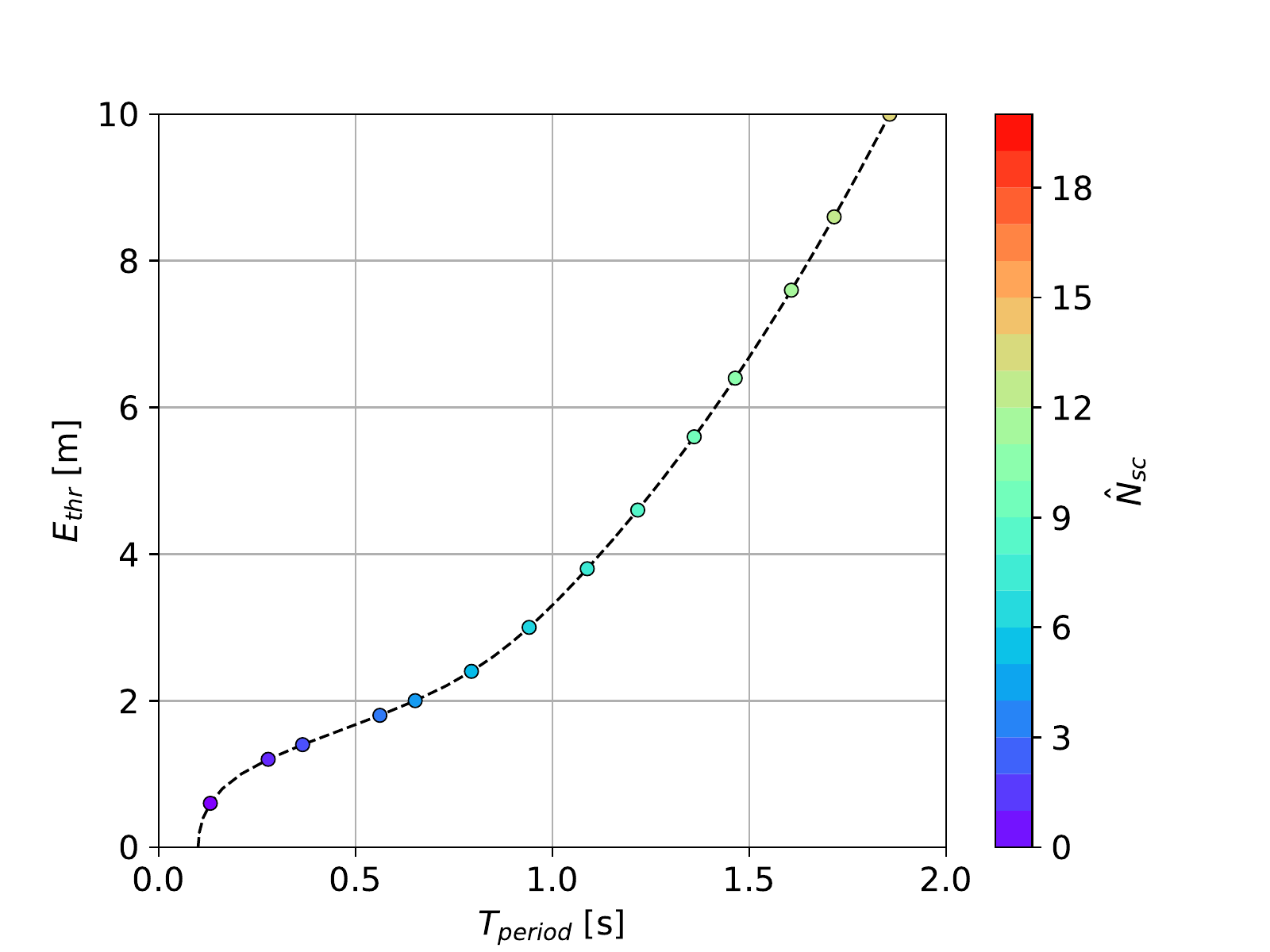}
	\caption{Function $\mathcal{F}^{-1}$.}
	\label{fig:intertx_error}
\end{figure}

In Fig.~\ref{fig:collision_prob}, we plot $P_{coll}$ as a function of $\rho$ for different values of $\hat{N}_{sc}=\lceil \frac{\hat N+1}{n_{sc}}\rceil$, which represents the estimate of the number of users  in the communication range of the ego vehicle  that use the same subcarrier of the ego vehicle itself.
By looking at Fig.~\ref{fig:collision_prob}, we observe that $P_{coll}$ increases with both the transmission probability and the number of interfering neighbors.
When using the \gls{nacc} approach, the ego vehicle sets the transmission probability $\rho$ according to both $\hat{N}_{sc}$ and $P_{coll}$.
Hence, the value of $T_{period}$ is updated as $\frac{1}{\rho}$ while the value of $E_{thr}$ is updated as $\mathcal{F}^{-1} \left(\frac{1}{\rho}\right)$.
In Fig.~\ref{fig:intertx_error}, we represent the function $\mathcal{F}^{-1}$ used for this purpose.
In particular, the colored dots  represent the different values of $E_{thr}$ that are chosen according to both $\hat{N}_{sc}$ and $P_{coll}$.

\begin{figure}[h!]
	\centering
	\includegraphics[width=4in]{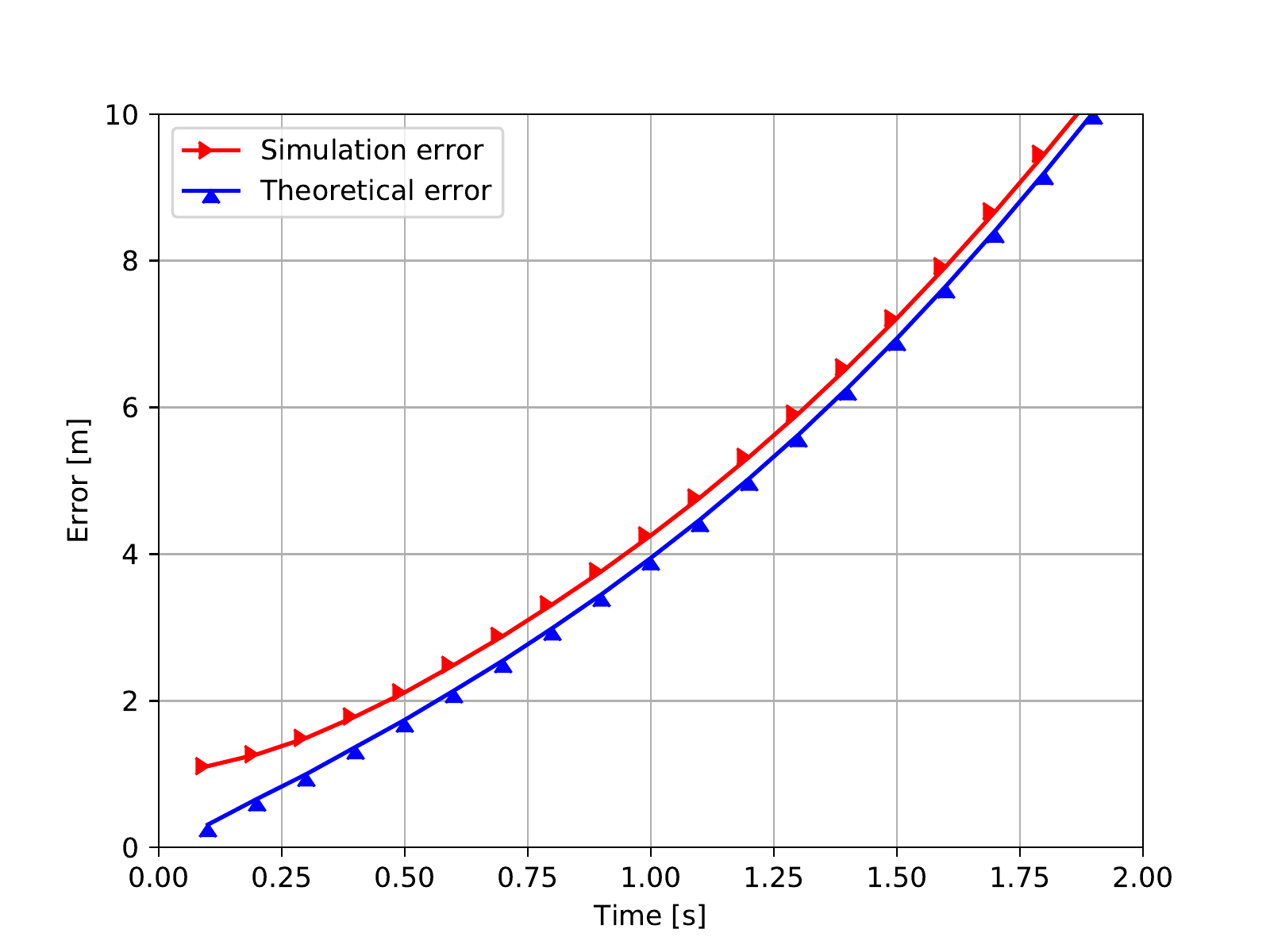}
	\caption{Simulation vs. theoretical positioning error in a  highway scenario.}
	\label{fig:validation}
\end{figure}

To validate our analysis, we show how the average tracking error, i.e., $E(E_{thr})$, evolves considering the output of the Kalman filter and the empirical results obtained through simulation.
In the first case, the motion of a vehicle is analytically represented as a rectilinear motion, disregarding the users' driving imperfections, and the vehicle's speed and direction are assumed constant. In the second case, mobility traces are generated using SUMO. In Fig.~\ref{fig:validation}, we represent the average positioning error obtained in the two different cases as a function of time. We observe that the two data trends are very similar, thereby validating our theoretical framework. The gap between the two curves is due to driving imperfections which cannot be predicted  by the  rectilinear motion model.

\subsection{Simulation Results}
\label{sub:sim_results}

\begin{figure}[t]
	\centering
	\includegraphics[width=4in]{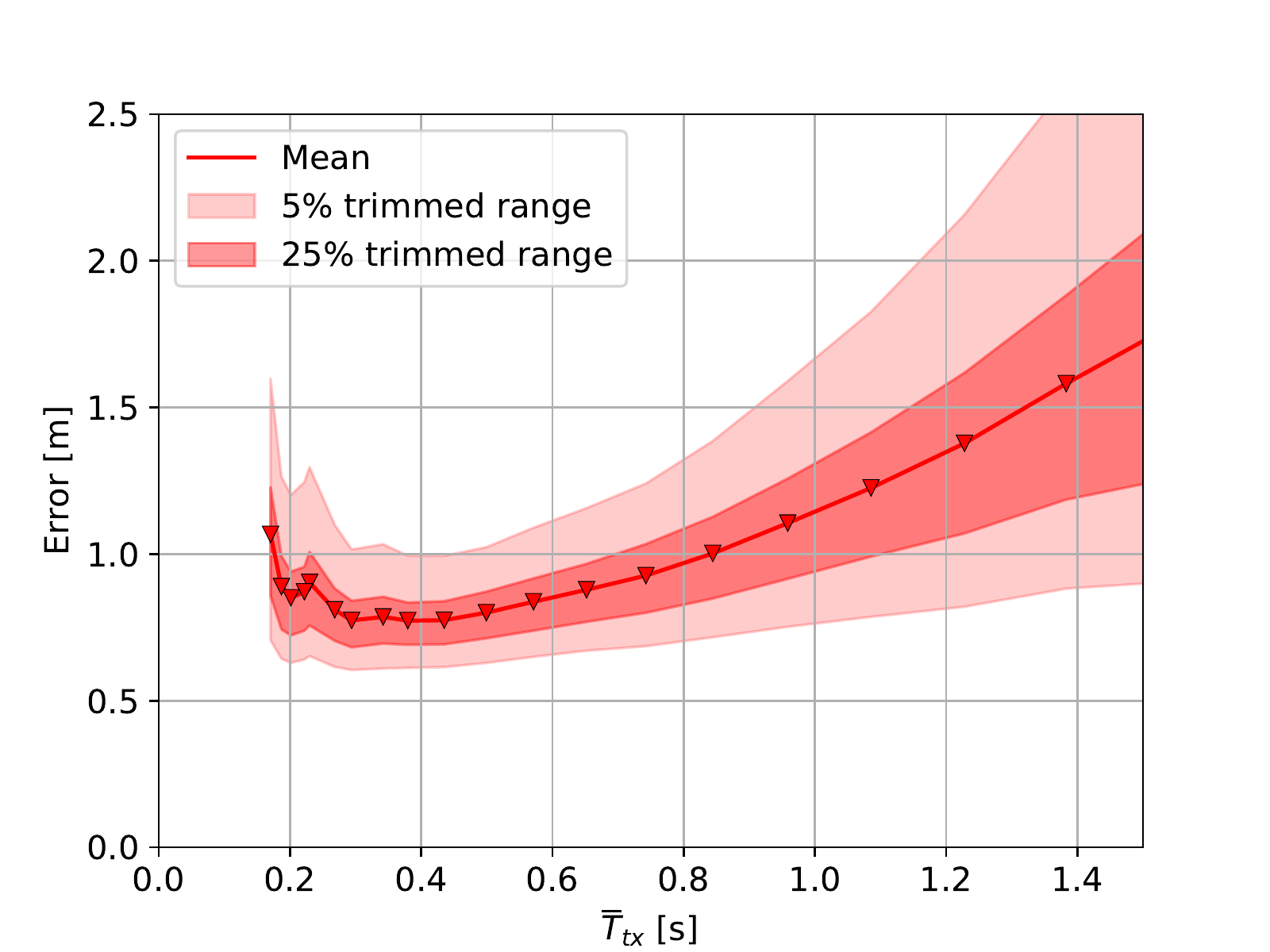}
	\caption{Positioning error statistics as a function of $\overline{T}_{tx}$, in case of $n_{sc}=8$ and \gls{pb} strategy.}
	\label{fig:error_policy0}
\end{figure}

\begin{figure}[t]
	\centering
	\includegraphics[width=4in]{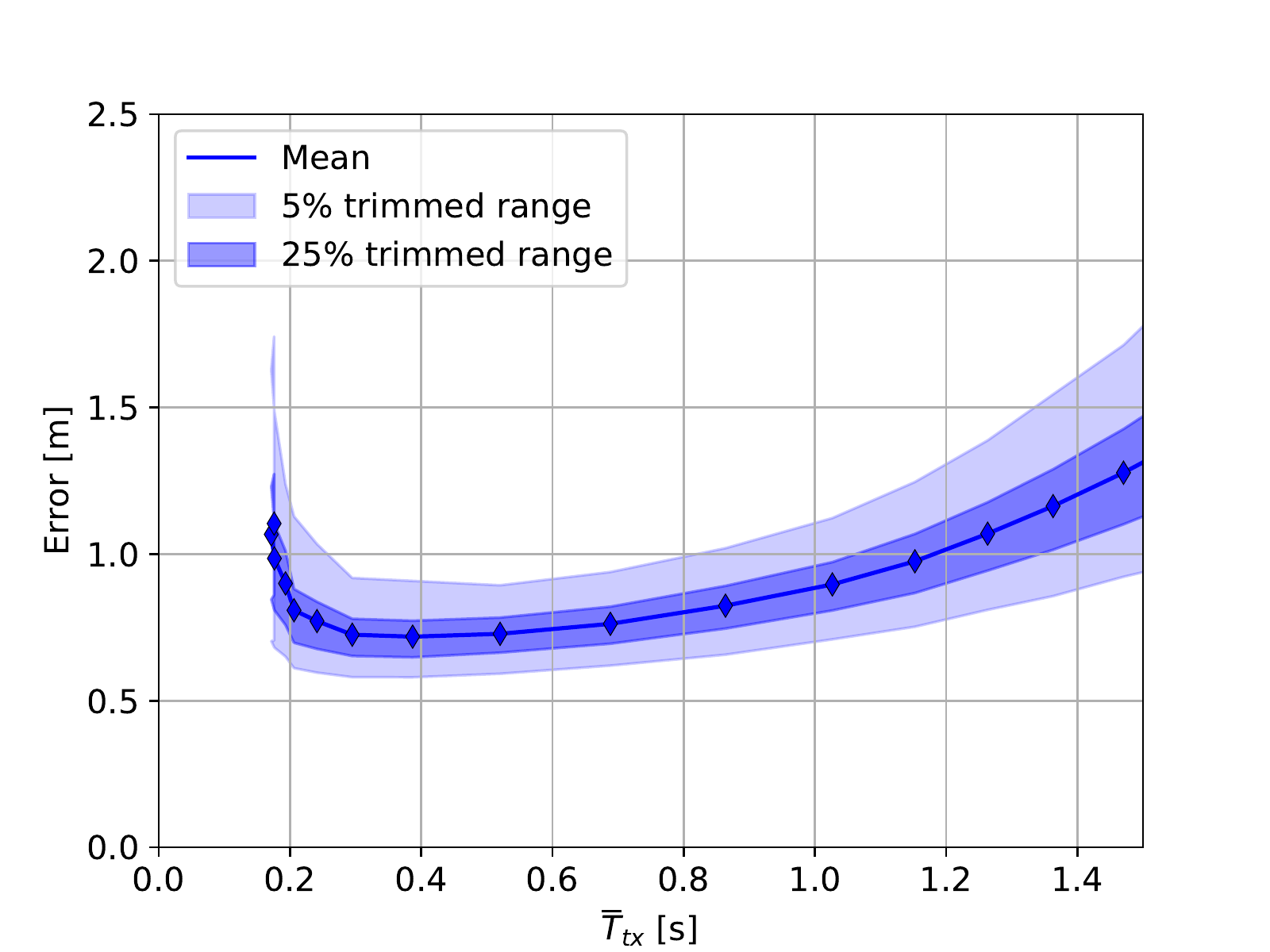}
	\caption{Positioning error statistics as a function of $\overline{T}_{tx}$, in case of $n_{sc}=8$ and \gls{etb} strategy.}
	\label{fig:error_policy1}
\end{figure}

In the rest of the section, we  analyze the performance of the broadcasting strategies and the congestion control schemes that we described in Sec.~\ref{sec:comm_strategy} and Sec.~\ref{sec:cong_con}, respectively. At first, we fix the number of the available subcarriers to $n_{sc}=8$. Later, we will verify how different $n_{sc}$ values may influence the simulation outcomes.
As we already stated, if we do not implement a congestion control mechanism, we have to determine \textit{a priori} the inner setting of the \gls{pb}  and the \gls{etb} strategies.
To fairly compare the performance of the two techniques, we adopt an exhaustive approach, obtaining a different outcome for each choice of $T_{period}$ and $E_{thr}$. In Fig.~\ref{fig:error_policy0} and Fig.~\ref{fig:error_policy1}, we analyze the statistics of the positioning error according to the mean inter-transmission time $\overline{T}_{tx}$, which is an indicator of the total channel occupancy. 


We highlight that $\overline{T}_{tx}$ does not coincide with the inter-transmission period used in the \gls{pb} strategy. Indeed, while $T_{period}$ is defined \textit{a priori} and can assume all the values within the  set $\{0.1\ \text{s},\text{...}, 10\ \text{s}\}$, $\overline{T}_{tx}$ is an outcome of the simulation.
In particular, in a realistic scenario,
$\overline{T}_{tx}$ never goes below the value of $ 0.2 $ s, i.e., two timeslots, because of the channel access contention.
From Fig.~\ref{fig:error_policy0} and Fig.~\ref{fig:error_policy1}, we can also observe how the limits of the \gls{csmaca} affect the positioning error: when the number of channel access requests is too high, i.e., $\overline{T}_{tx} < 0.3$ s, the channel gets congested and, consequently, the performance of the overall system degrades.
Indeed, the positioning error can be described by a convex curve, with a minimum for $\overline{T}_{tx} \approx 0.3$ s; this value represents the level of channel occupancy that guarantees the best position estimation accuracy.
By comparing Fig.~\ref{fig:error_policy0} with Fig.~\ref{fig:error_policy1}, we can observe that the \gls{etb} strategy outperforms the benchmark \gls{pb} strategy. In particular, \gls{etb} proves to have a slightly lower average error and a significantly lower error variance.

\begin{figure}[t]
	\centering
	\includegraphics[width=4in]{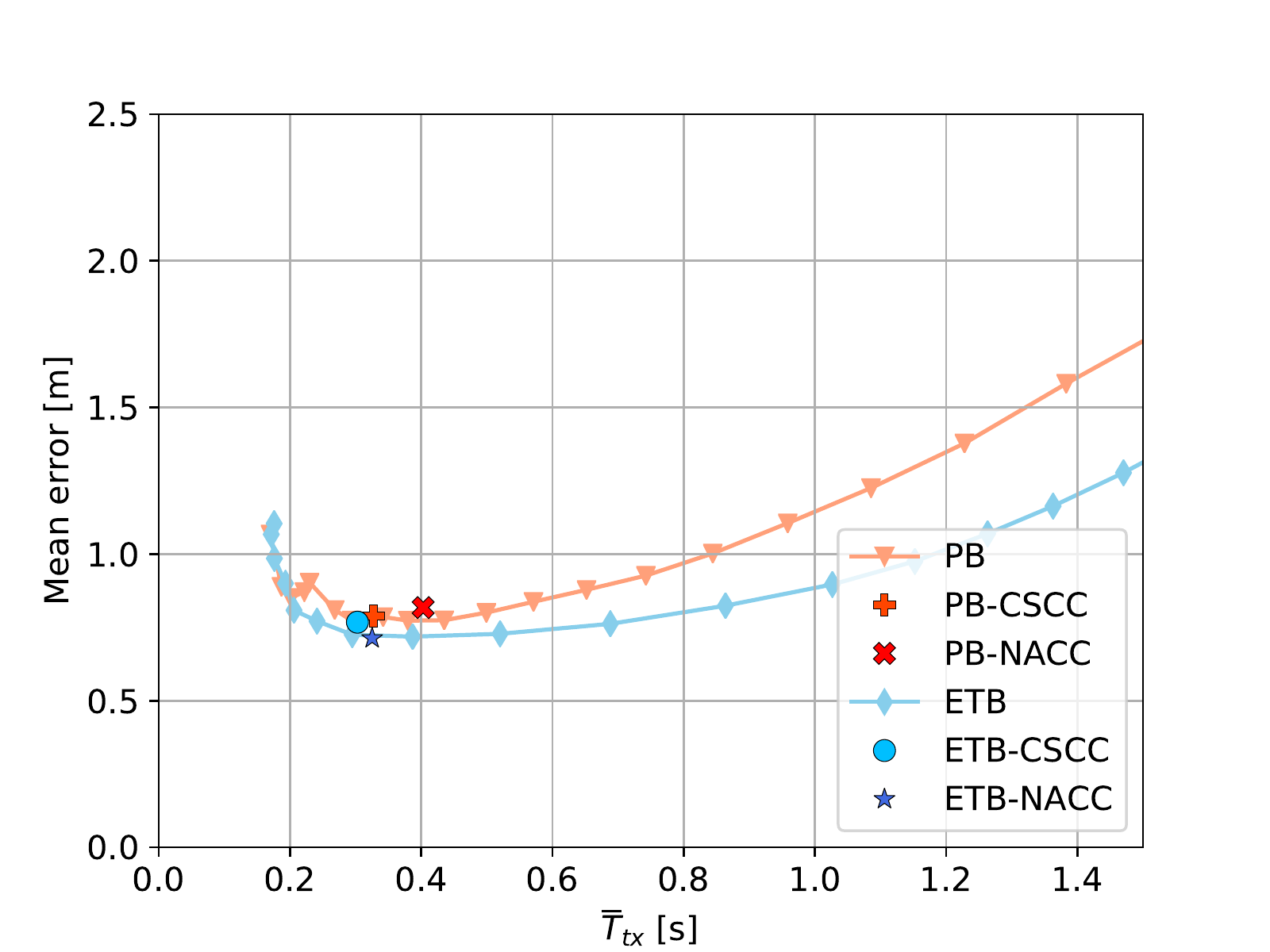}
	\caption{Mean positioning error as a function of $\overline{T}_{tx}$, in case of $n_{sc}=8$.}
	\label{fig:avgerror_exhaustive}
\end{figure} 
\begin{figure}[t]
	\centering
	\includegraphics[width=4in]{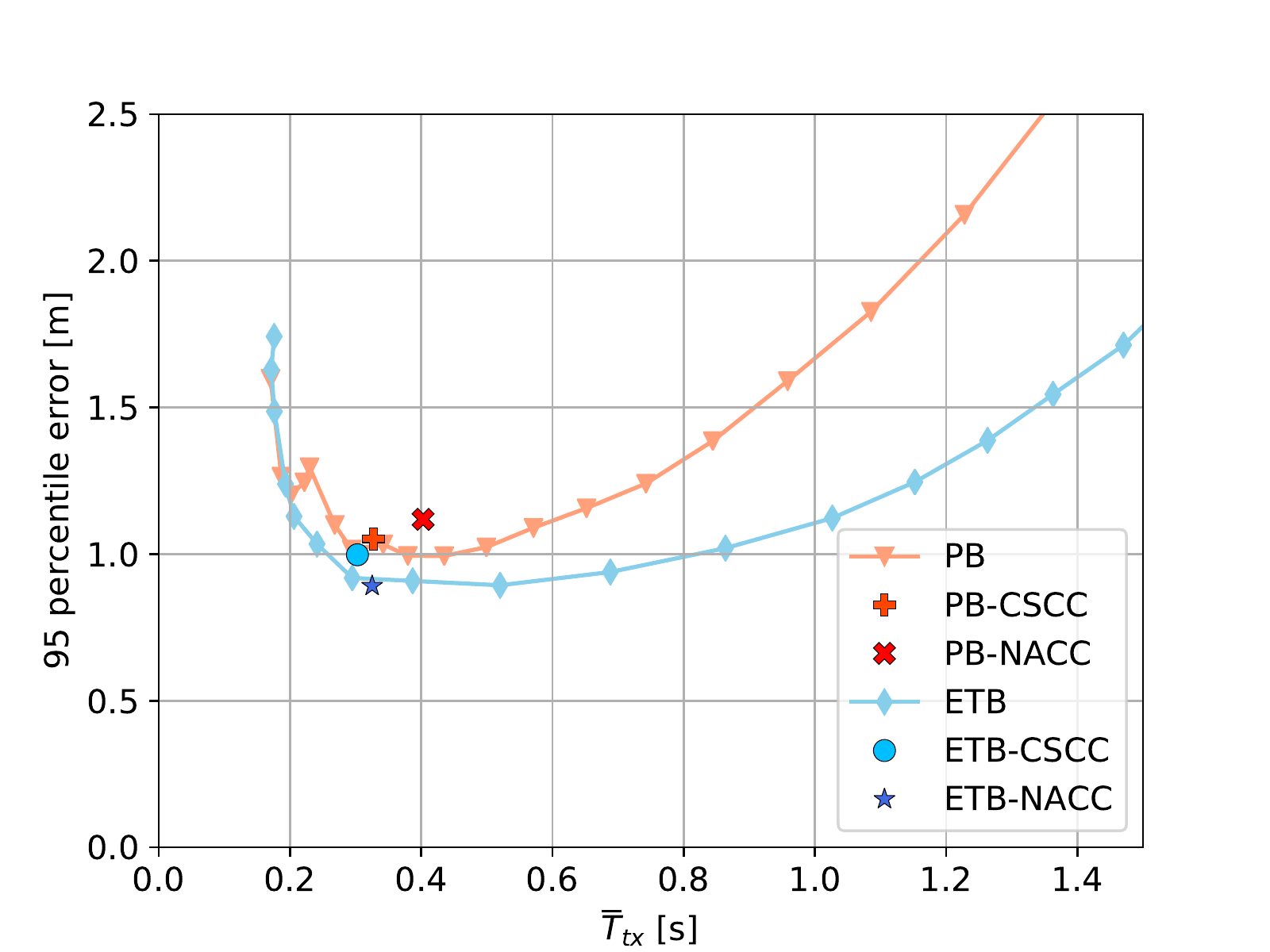}
	\caption{95 percentile of the positioning error as a function of $\overline{T}_{tx}$, in case of $n_{sc}=8$.}
	\label{fig:95error_exhaustive}
\end{figure}

By optimizing the channel access requests, the \gls{etb} strategy has a lower positioning error than the benchmark strategy.
Fig.~\ref{fig:avgerror_exhaustive} and Fig.~\ref{fig:95error_exhaustive} shows a direct comparison between the considered broadcasting strategies; in particular, Fig.~\ref{fig:avgerror_exhaustive} reports the mean error while Fig.~\ref{fig:95error_exhaustive} shows the 95th percentile of the error.
In both cases, the \gls{etb} strategy ensures better position estimation accuracy for the same level of channel occupancy.
The marks in Fig.~\ref{fig:avgerror_exhaustive} and Fig.~\ref{fig:95error_exhaustive} represent the performance of the congestion control schemes designed in Sec.\ref{sec:cong_con}.\footnote{Since congestion control can adapt the communication strategy to the scenario in real-time, we obtain a single outcome for each combination of broadcasting strategy and congestion control approach.}
First, we observe that all the deployed solutions succeed in maintaining the channel occupancy close to the optimal working point, i.e., $\overline{T}_{tx} \approx 0.3$ s. 
Among all the possible solutions, the combination of the \gls{etb} strategy with the \gls{nacc} approach ensures the best performance.
In particular, this scheme outperforms the classical approach used in the literature, which is represented by the combination of the \gls{pb} strategy with the \gls{cscc} approach, obtaining a $10\%$ gain when considering the mean error and $20\%$ gain when considering the 95th percentile of the error.

\begin{figure}[t!]
	\centering
	\setlength{\belowcaptionskip}{-5pt}
	\includegraphics[width=4in]{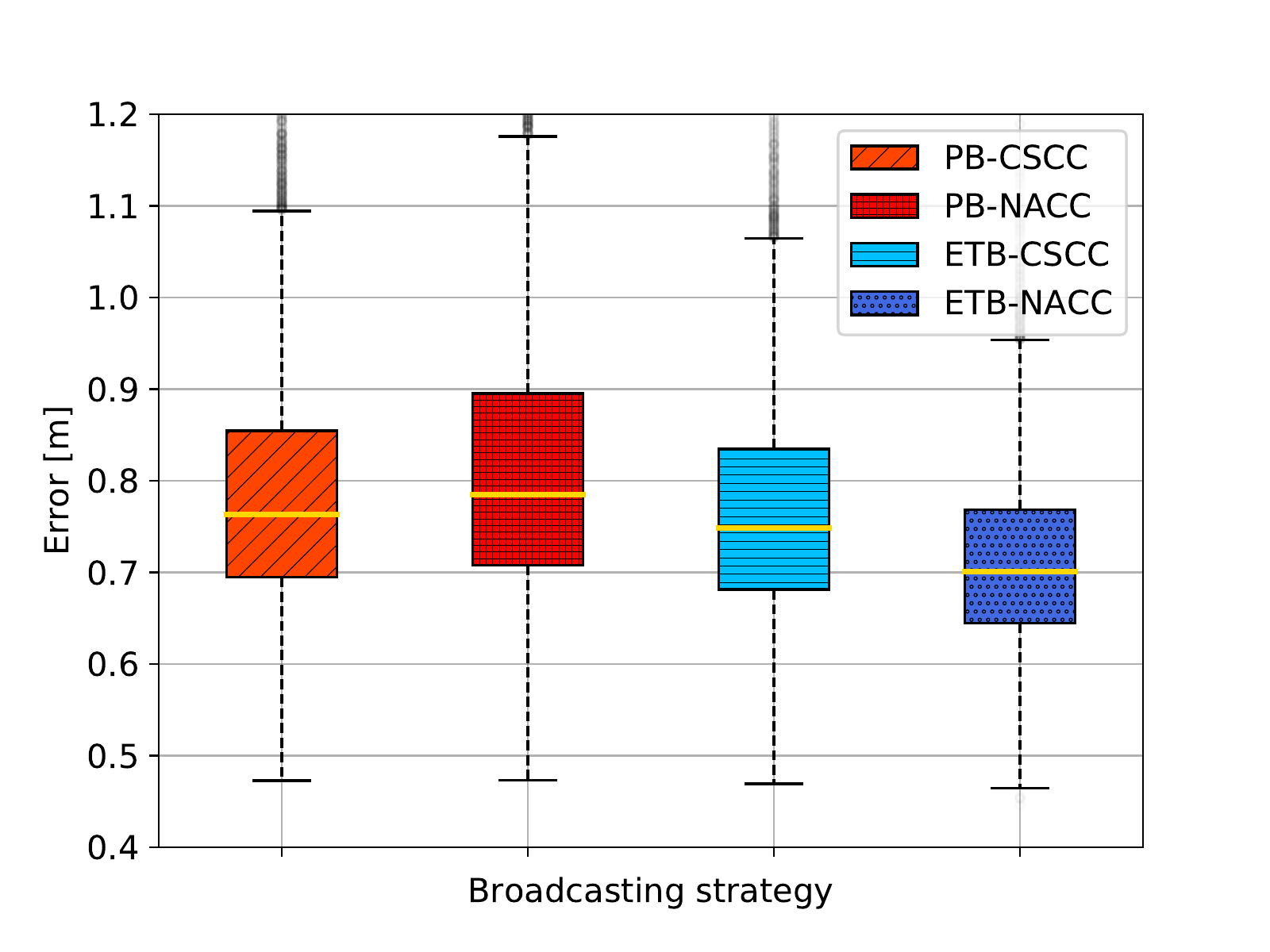}
	\caption{Boxplot of the positioning error with $n_{sc}=8$.}
	\label{fig:error_box}
\end{figure}
The full positioning error statistics of the four congestion control solutions are shown as a boxplot in Fig.~\ref{fig:error_box}.
In the figure, each box is delimited by the first and the third quartiles of the error distribution. The box's center lines represent the median of the error, and the whiskers show the 95\% confidence intervals. Outliers are represented as dots. We can see that our solution  is the only technique that ensures that the third quartile is below $0.8$ m and that the confidence interval is below $1.0$ m. 

\begin{figure}[t!]
	\centering
	\begin{subfigure}[t]{0.44\columnwidth}
		\centering
		\includegraphics[width=3in]{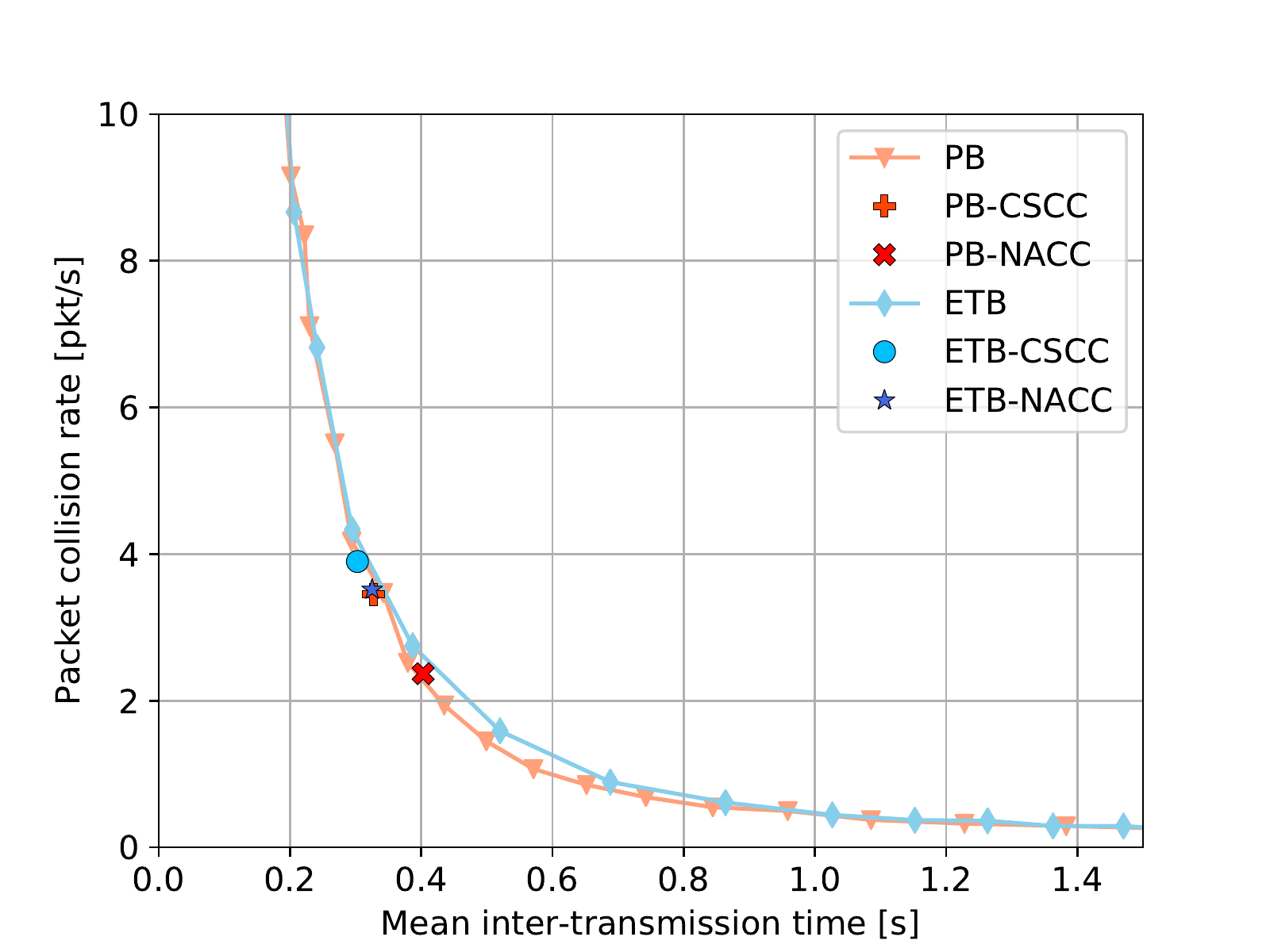}
		\caption{Packet collision rate.}
		\label{fig:collision_exhaustive}
	\end{subfigure} 
	\begin{subfigure}[t]{0.44\columnwidth}
		\centering
		\includegraphics[width=3in]{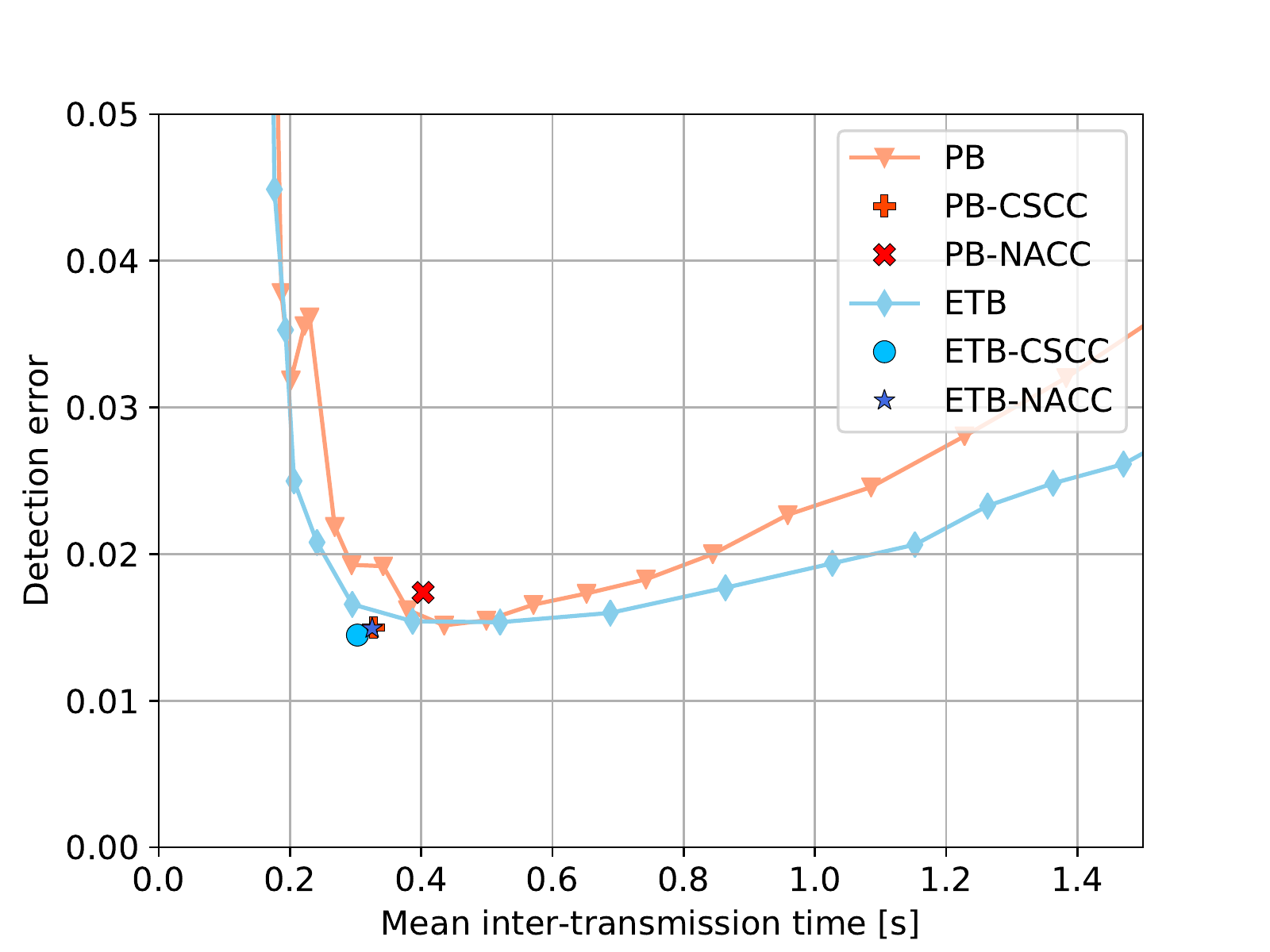}
		\caption{Detection error probability.}
		\label{fig:detection_exhaustive}
	\end{subfigure}
	\caption{Collision and detection statistics as a function of the average inter-transmission time, with $n_{sc}=8$.}
	\label{fig:coll_det_exhaustive}
\end{figure}

In Fig.~\ref{fig:coll_det_exhaustive} we show the packet collision rate and the detection error probability.
Taking into account the broadcasting strategies without the congestion control schemes, we observe that both techniques present almost identical trends.
As we can observe from Fig.~\ref{fig:collision_exhaustive}, the amount of information that gets lost in the channel significantly increases when $\overline{T}_{tx} < 0.5$ s, independently of the deployed strategy.
This phenomenon explains the degradation of the positioning estimation accuracy that we observe in Fig.~\ref{fig:error_policy0} and Fig.~\ref{fig:error_policy1}.
We highlight that the channel congestion does not affect only the positioning error but also the probability of misdetection and false alarm of a neighbor vehicle. Indeed, by looking at Fig.~\ref{fig:detection_exhaustive}, we can observe that the detection error probability increases exponentially as soon as $\overline{T}_{tx} < 0.25$ s.
Since the strategies' optimal working point is $\overline{T}_{tx} \approx 0.3$, we conclude that minimizing the positioning error does not necessarily imply an increase of the detection error probability.
Considering the congestion control approaches, we observe that none of the obtained outcomes deviate from the curves defined by the exhaustive simulations.
As already mentioned, the combination of the \gls{pb} strategy with the \gls{nacc} approach presents a slightly higher $\overline{T}_{tx}$ and, therefore, is characterized by a different packet collision rate and detection error probability. 

\begin{figure}[t]
		\centering
		\includegraphics[width=4in]{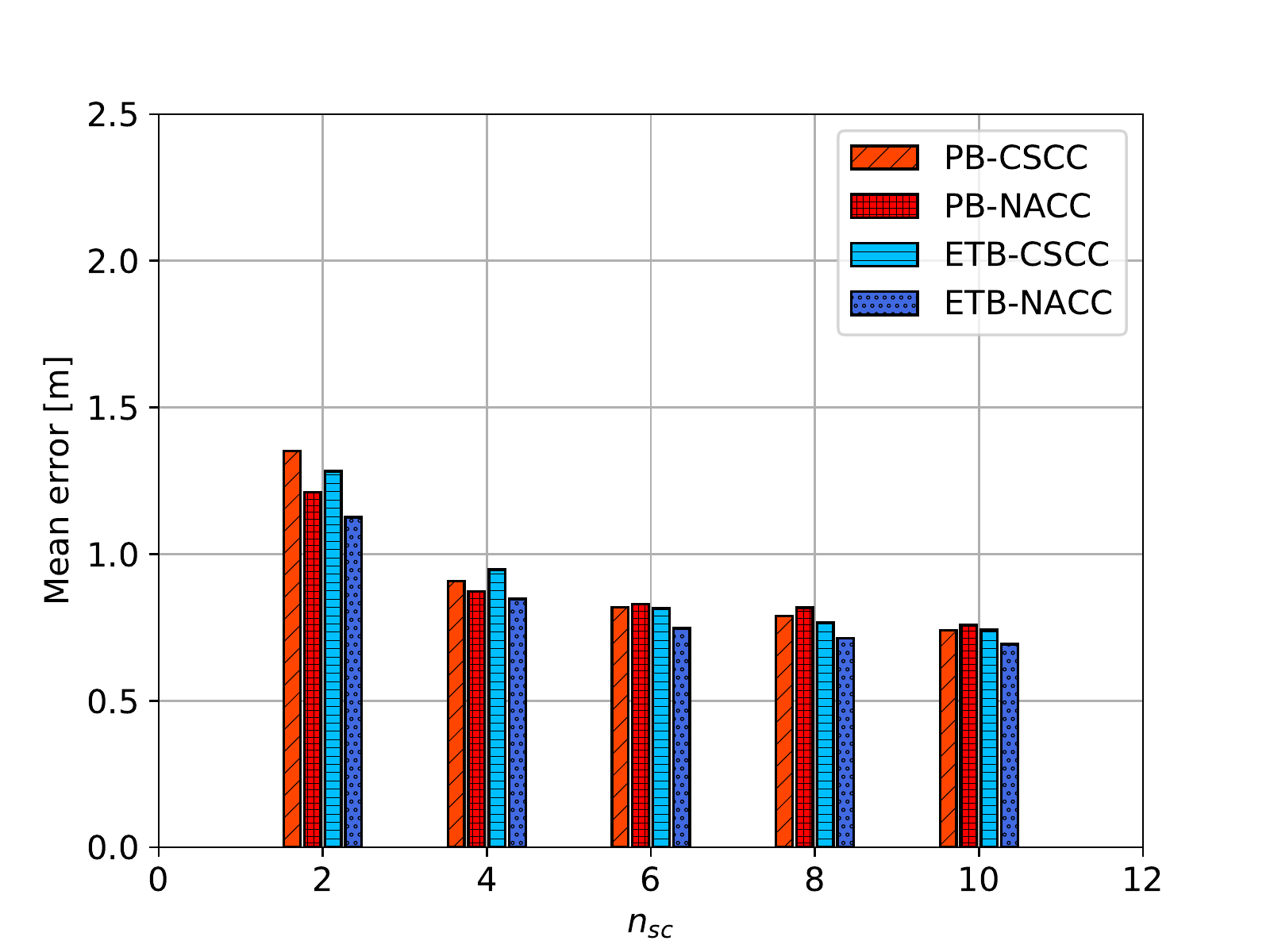}
		\caption{Mean positioning error as a function of $n_{sc}$.}
		\label{fig:error_bar}
\end{figure}
 
\begin{figure}[t]
		\centering
		\includegraphics[width=4in]{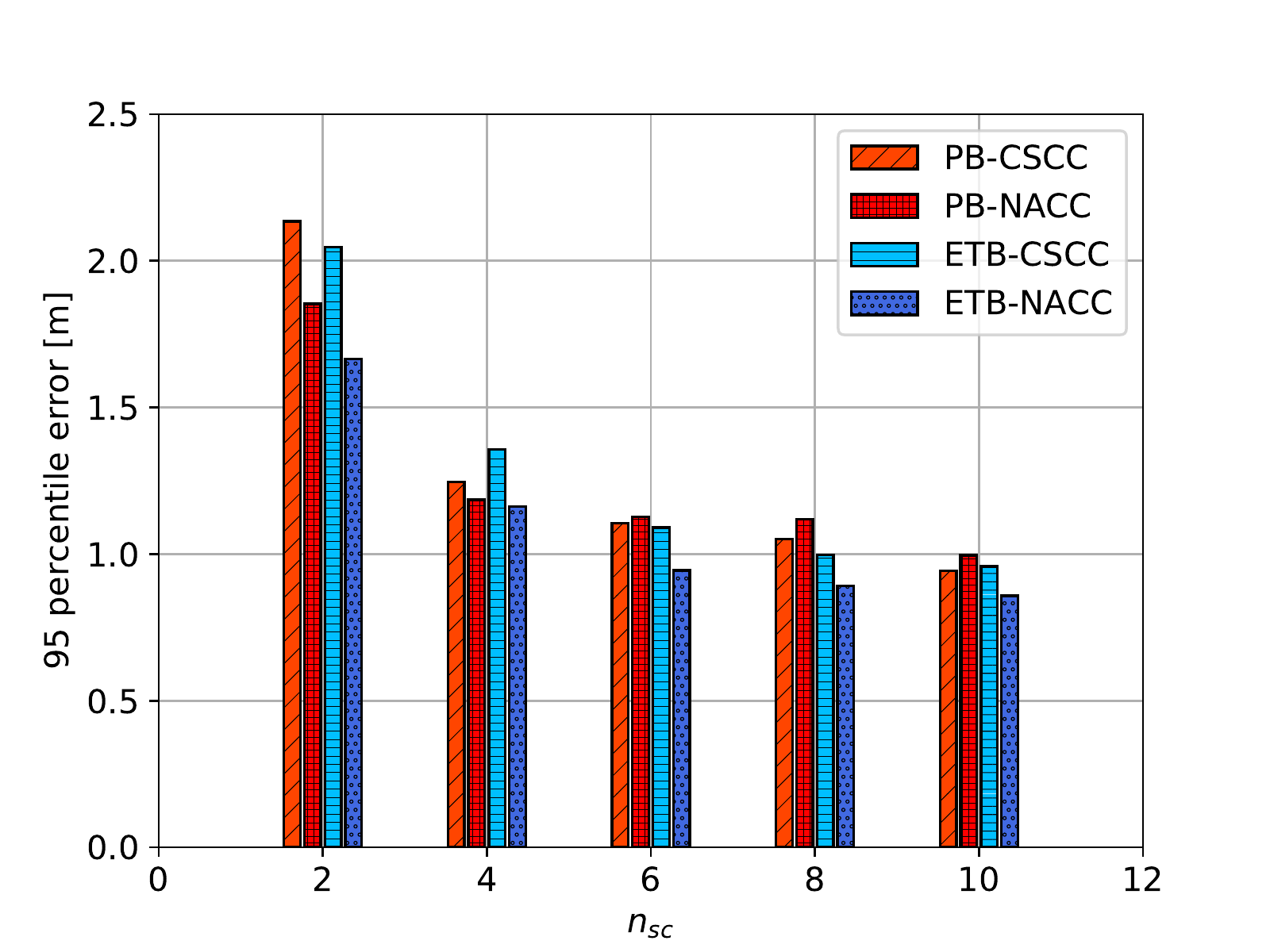}
		\caption{95th percentile of the positioning error as a function of $n_{sc}$.}
		\label{fig:95error_bar}
\end{figure}

In order to validate these results in a more general scenario, we analyze the performance of the four possible congestion control schemes with a different numbers of subcarriers $n_{sc}$.
The results of this analysis are reported in Fig.~\ref{fig:error_bar} and Fig.~\ref{fig:95error_bar}.
We observe that the solution combining the \gls{etb} strategy and the \gls{nacc} approach outperforms the other schemes for any value of $n_{sc}$, considering both the mean positioning error (in Fig.~\ref{fig:error_bar}) and the 95th percentile of the positioning error (in Fig.~\ref{fig:95error_bar}).
In particular, our solution outperforms state of the art solutions by up to $20\%$ mean error reduction and up to $30\%$ 95th percentile error reduction.
For what concerns the other techniques, we observe that the combination of \gls{etb} and \gls{cscc} performs poorly for $n_{sc}\leq 4$, while it leads to better results when the number of subcarriers is greater.
Conversely, the combination of \gls{pb} and \gls{nacc} performs well for $n_{sc}\leq4$ but does not fully exploit the available resources when $n_{sc}\geq6$.

\vspace{-8pt}
\section{Conclusions and Future Work}
\label{sec:conclusions}
 
In this work, we  studied the trade-off between ensuring accurate position information and preventing congestion of the communication channel in vehicular networks and designed an innovative threshold-based broadcasting algorithm  that forces vehicles to distribute state information if the estimated positioning error is above a certain error threshold. We also adopted a new congestion control mechanism that adapts the inter-transmission period according to network topology information.
We showed through simulations that the proposed approach outperforms a conventional broadcasting strategy, which relies on a periodic transmission of state information and channel sensing, since it reduces the positioning error with no additional resources.

As part of our future work, we will test our broadcasting and congestion control frameworks in more complex scenarios, e.g., by considering different road maps and traffic conditions. Besides, we are interested in improving the tracking accuracy of our communication strategy by exploiting the \gls{ml} paradigm to take into account the results of previous broadcasting operations as a bias to decide whether and when to transmit state information updates.

\ifCLASSOPTIONcaptionsoff
 \newpage
\fi



\bibliographystyle{IEEEtran}
\vspace{-10pt}
\bibliography{IEEEabrv,bibliography}

%








\end{document}